\documentclass[ALICE,manyauthors]{cernphprep}

\usepackage[comma,square,numbers,sort&compress]{natbib}
\usepackage{hyperref}
\usepackage{booktabs}
\usepackage{lineno}
\usepackage{color}
\usepackage{multirow}
\usepackage[T1]{fontenc}
\usepackage{textcomp}
\newcommand{\ppip}{\ensuremath{\pi^{+}}\xspace}
\newcommand{\ppim}{\ensuremath{\pi^{-}}\xspace}
\newcommand{\ppi}{\ensuremath{\pi}\xspace}
\newcommand{\ppipm}{\ensuremath{\pi^{\pm}}\xspace}
\newcommand{\ppis}{\ensuremath{\ppip + \ppim}\xspace}
\newcommand{\pkap}{\ensuremath{{\rm K}^{+}}\xspace}
\newcommand{\pkam}{\ensuremath{{\rm K}^{-}}\xspace}
\newcommand{\pka}{\ensuremath{{\rm K}}\xspace}
\newcommand{\pkapm}{\ensuremath{{\rm K}^{\pm}}\xspace}
\newcommand{\pkas}{\ensuremath{\pkap + \pkam}\xspace}

\newcommand{\pkazero}{\ensuremath{{\rm K}^{0}_{\rm S}}\xspace}
\newcommand{\pprp}{\ensuremath{{\rm p}}\xspace}
\newcommand{\pprm}{\ensuremath{\overline{\rm{p}}}\xspace}
\newcommand{\ppr}{\ensuremath{{\rm p}}\xspace}

\newcommand{\pprs}{\ensuremath{\pprp + \pprm}\xspace}
\newcommand{\pphi}{\ensuremath{\phi}\xspace}
\newcommand{\plam}{\ensuremath{\Lambda}\xspace}
\newcommand{\psigp}{\ensuremath{\Sigma^{+}}\xspace}
\newcommand{\rppi}{\ensuremath{\ppr/\ppi}\xspace}
\newcommand{\rphipi}{\ensuremath{\pphi/\ppi}\xspace}
\newcommand{\rpphi}{\ensuremath{\ppr/\pphi}\xspace}
\newcommand{\rkpi}{\ensuremath{\pka/\ppi}\xspace}

\newcommand{\ndfITS}{\ensuremath{N^{\rm hits}_{\rm ITS}}\xspace}
\newcommand{\chitwo}{\ensuremath{\chi^{2}}\xspace}
\newcommand{\chitwoITS}{\ensuremath{\chi^{2}_{\rm ITS}}\xspace}
\newcommand{\chitwoNDF}{\ensuremath{\chi^{2}/{\rm NDF}}\xspace}
\newcommand{\chitwoCluster}{\ensuremath{\chi^{2}/{\rm cluster}}\xspace}
\newcommand{\dca}{\ensuremath{{\rm DCA}}\xspace}
\newcommand{\dcaXY}{\ensuremath{\dca_{\it{xy}}}\xspace}
\newcommand{\sigmadcaXY}{\ensuremath{\sigma_{\dcaXY}}\xspace}
\newcommand{\dcaZ}{\ensuremath{\dca_{\it{z}}}\xspace}
\newcommand{\mom}{\ensuremath{p}\xspace}
\newcommand{\momTRUE}{\ensuremath{\mom^{\rm TRUE}}\xspace}
\newcommand{\momMEAS}{\ensuremath{\mom^{\rm MEAS}}\xspace}
\newcommand{\pt}{\ensuremath{p_{\rm{T}}}\xspace}
\newcommand{\ptTRUE}{\ensuremath{\pt^{\rm TRUE}}\xspace}
\newcommand{\ptMEAS}{\ensuremath{\pt^{\rm MEAS}}\xspace}
\newcommand{\meanpt}{\ensuremath{\langle \pt \rangle}\xspace}
\newcommand{\mt}{\ensuremath{m_{\rm{T}}}\xspace}
\newcommand{\qt}{\ensuremath{q_{\rm{T}}}\xspace}
\newcommand{\cmc}{\ensuremath{\text{cm}/c}\xspace}
\newcommand{\Nsigma}{\ensuremath{N_{\sigma}}\xspace}
\newcommand{\nsigma}{\ensuremath{\Nsigma}\xspace}
\newcommand{\NsigmaTPC}{\ensuremath{\Nsigma^{\rm TPC}}\xspace}
\newcommand{\NsigmaTOF}{\ensuremath{\Nsigma^{\rm TOF}}\xspace}
\newcommand{\abs}[1]{\ensuremath{\vert #1 \vert}\xspace}
\newcommand{\NsigmaTPCabs}{\ensuremath{\abs{\NsigmaTPC}}\xspace}
\newcommand{\NsigmaTOFabs}{\ensuremath{\abs{\NsigmaTOF}}\xspace}
\newcommand{\tofmeas}{\ensuremath{\text{time-of-flight}}\xspace}
\newcommand{\AAaa}{\ensuremath{{\rm{AA}}}\xspace}

\newcommand{\dedx}{\ensuremath{{\rm d}E/{\rm d}x}\xspace}

\newcommand{\dndy}{\ensuremath{{\rm d}N/{\rm d}y}\xspace}
\newcommand{\avdndy}{\ensuremath{ \langle \dndy \rangle }\xspace}
\newcommand{\avdndypart}[1]{\ensuremath{\avdndy_{#1}}\xspace}
\newcommand{\Nch}{\ensuremath{{\it N}_{\rm{ch}}}\xspace}
\newcommand{\dNchdeta}{\ensuremath{\rm{d}\Nch/\rm{d}\eta}\xspace}
\newcommand{\avdNchdeta}{\ensuremath{\langle\dNchdeta\rangle}\xspace}

\newcommand{\EcrossB}{\ensuremath{E\times B}\xspace}
\newcommand{\interval}[2]{\ensuremath{#1{-}#2}\xspace}
\newcommand{\centint}[2]{\ensuremath{\interval{#1}{#2}\%}\xspace}
\newcommand{\momint}[2]{\ensuremath{\interval{#1}{#2}~\gevc}\xspace}
\newcommand{\minv}{\ensuremath{M_{\rm{KK}}}\xspace}
\newcommand{\vtwo}{\ensuremath{v_{2}}\xspace}
\newcommand{\vtwotwo}{\ensuremath{\vtwo\{2, \abs{\Delta\eta}>2\}}\xspace}
\newcommand{\degrees}[1]{\ensuremath{#1^{\rm o}}\xspace}

\newcommand{\xexe}{\ensuremath{\text{Xe--Xe}}\xspace}
\newcommand{\pbpb}{\ensuremath{\text{Pb--Pb}}\xspace}
\newcommand{\ppb}{\ensuremath{\text{p--Pb}}\xspace}
\newcommand{\pp}{\ensuremath{\text{pp}}\xspace}
\newcommand{\gevc}{\ensuremath{{\rm GeV}/c}\xspace}
\newcommand{\mevc}{\ensuremath{{\rm MeV}/c}\xspace}
\newcommand{\gevcsq}{\ensuremath{{\rm GeV}/c^{2}}\xspace}
\newcommand{\mevcsq}{\ensuremath{{\rm MeV}/c^{2}}\xspace}

\newcommand{\s}{\ensuremath{\sqrt{s}}\xspace}
\newcommand{\sF}{\ensuremath{\s~=~5.02~\text{TeV}}\xspace}
\newcommand{\snn}{\ensuremath{\sqrt{s_{\rm{NN}}}}\xspace}
\newcommand{\snnT}{\ensuremath{\snn~=~2.76~\text{TeV}}\xspace}
\newcommand{\snnF}{\ensuremath{\snn~=~5.02~\text{TeV}}\xspace}
\newcommand{\snnXeXe}{\ensuremath{\snn~=~5.44~\text{TeV}}\xspace}

\ifdefined\Ref
  \renewcommand{\Ref}[1]{Ref.\cite{#1}\xspace}
\else
  \newcommand{\Ref}[1]{Ref.\cite{#1}\xspace}
\fi

\newcommand{\Section}[1]{Section~\ref{#1}\xspace}
\newcommand{\Tab}[1]{Table~\ref{#1}\xspace}
\newcommand{\Eq}[1]{Eq.~\ref{#1}\xspace}

\newcommand{\Fig}[1]{Fig.~\ref{#1}\xspace}

\newcommand{\BR}{B.R.\xspace}

\newcommand{\GEANTT}{\ensuremath{\text{GEANT3}}\xspace}
\newcommand{\GEANTF}{\ensuremath{\text{GEANT4}}\xspace}
\newcommand{\FLUKA}{\ensuremath{\text{FLUKA}}\xspace}
\newcommand{\HIJING}{\ensuremath{\text{HIJING}}\xspace}

\newcommand{\ALICE}{\ensuremath{\text{ALICE}}\xspace}

\begin{document}%

\begin{titlepage}
  \PHyear{2020}
  \PHnumber{249}      
  \PHdate{22 December}  
  %

  \title{Production of pions, kaons, (anti-)protons and \pphi mesons in \xexe collisions at \textbf{\snn}~=~5.44 TeV}
  \ShortTitle{Production of \ppipm, \pkapm, \pprp, \pprm and \pphi in \xexe collisions at \textbf{\snn}~=~5.44 TeV}   
  \Collaboration{ALICE Collaboration\thanks{See Appendix~\ref{app:collab} for the list of collaboration members}}
  \ShortAuthor{ALICE Collaboration} 

  The first measurement of the production of pions, kaons, (anti-)protons and \pphi mesons at midrapidity in \xexe collisions at \snnXeXe is presented.
Transverse momentum (\pt) spectra and \pt-integrated yields are extracted in several centrality intervals bridging from \ppb to mid-central \pbpb collisions in terms of final-state multiplicity.
The study of \xexe and \pbpb collisions allows systems at similar charged-particle multiplicities but with different initial geometrical eccentricities to be investigated.
A detailed comparison of the spectral shapes in the two systems reveals an opposite behaviour for radial and elliptic flow.
In particular, this study shows that the radial flow does not depend on the colliding system when compared at similar charged-particle multiplicity.
In terms of hadron chemistry, the previously observed smooth evolution of particle ratios with multiplicity from small to large collision systems is also found to hold in \xexe.
In addition, our results confirm that two remarkable features of particle production at LHC energies are also valid in the collision of medium-sized nuclei: the lower proton-to-pion ratio with respect to the thermal model expectations and the increase of the \pphi-to-pion ratio with increasing final-state multiplicity.

\end{titlepage}
\setcounter{page}{2}

\section{Introduction} \label{sec:intro}
In recent years, the production of hadrons consisting of light flavour quarks ($u$, $d$, and $s$) has been extensively studied in pp, \ppb and \pbpb collisions at LHC energies~\cite{Abelev:2013haa,Abelev:2013vea,Adam:2015vsf,Aamodt:2011zza,Aamodt:2011zj,Adam:2016bpr,Adam:2017zbf,ALICE:2017jyt,Acharya:2018orn,Acharya:2019kyh, Acharya:2019qge} with the aim to explore the strongly interacting Quark-Gluon Plasma (QGP) produced in heavy-ion collisions.
After the formation, the QGP expands hydrodynamically reaching first a chemical freeze-out, where hadron abundances are fixed~\cite{Braun-Munzinger:2015hba, Adamczyk:2017iwn}, and then a kinetic freeze-out, where the hadron momenta are fixed. \\
Remarkably, a smooth evolution of the hadron chemistry, i.e. of the relative abundance of hadron species, was observed across different collision systems as a function of the final-state multiplicity~\cite{Acharya:2018orn}.
This behaviour was also found to be independent of collision energy~\cite{Acharya:2019kyh}.
In particular, the relative abundance of strange particles with respect to the non-strange ones increases continuously from small to large multiplicities until a saturation is observed for systems in which about 100 charged particles are produced per unit of pseudorapidity~\cite{ALICE:2017jyt}.
This observation suggests a gradual approach to a chemical equilibrium that is assumed to originate from the same underlying physical mechanisms across different collision systems~\cite{Kurkela:2018xxd,Bierlich:2018lbp,Bierlich:2014xba}.
The study of the pion, kaon, (anti-)proton, and \pphi production in the collisions of medium-sized nuclei such as Xe provides the ultimate test for validating this picture by bridging the gap between \ppb and \pbpb multiplicities.

In this context, two remarkable features of particle production are of particular interest to be verified in \xexe collisions: (i) the low value of the \rppi ratio with respect to statistical-thermal model estimates~\cite{Abelev:2012wca} and (ii) the rising trend of the \rphipi ratio from low to high multiplicities~\cite{Acharya:2018orn}.
The first observation has led to several speculations ranging from the incomplete treatment of resonance feed-down to a potential difference in chemical freeze-out temperatures among different quark flavours~\cite{Noronha-Hostler:2014aia,Bellwied:2013cta,Vovchenko:2018fmh} but found its most likely explanation in the inclusion of pion-nucleon phase shifts within the statistical-thermal model framework~\cite{Andronic:2018qqt}.
The second effect provides strict constraints for both the canonical statistical-thermal approach in which no rise is predicted~\cite{Acharya:2018orn,Vovchenko:2019kes,Sharma:2018jqf} as well as for models with only partial strangeness equilibration in which a steeper rise is expected similarly to the $\Xi$ baryon~\cite{Vovchenko:2019kes}.

Moreover, the detailed comparison of spectral shapes in \xexe and \pbpb collisions at similar multiplicities provides the unique opportunity to investigate the hydrodynamic expansion in systems of similar final state charged particle multiplicity and different geometrical eccentricity.
Already existing data on the elliptic flow coefficient \vtwo~\cite{Acharya:2018ihu} show a large difference in central collisions between the two systems, as expected from the Glauber and hydrodynamical models.
In contrast, the radial flow and consequently the mean transverse momenta are expected to be comparable between \xexe and \pbpb at similar multiplicities~\cite{Giacalone:2017dud}.
The test of this hypothesis is one of the subjects of this manuscript.
In addition, the data used in this article were collected with a lower magnetic field, thus allowing us to extend the measurement of pions to lower transverse momenta with respect to previous studies~\cite{PhysRevC.101.044907}.
For this reason, these data may also be of great relevance for future studies of potential condensation phenomena at low transverse momenta~\cite{Begun:2015ifa}.

This article is organised as follows.
\Section{Sec.:Analysis} describes the experimental setup and data analysis as well as the systematic uncertainties.
Results and comparisons with model calculations are discussed in \Section{Sec.:Results}.
The summary and conclusions are given in \Section{Sec.:Conclusions}.

\section{Experimental apparatus, data sample and analysis} \label{sec:analysis}
\label{Sec.:Analysis}
The measurements reported in this article are obtained with the \ALICE central barrel which has full azimuthal coverage around midrapidity in $|\eta|$ $<$ 0.8~\cite{Aamodt:2008zz}.
A detailed description of the full \ALICE apparatus can be found in~\cite{Abelev:2014ffa}.
In October 2017, for the first time at the LHC, \xexe collisions at \snnXeXe were recorded by the ALICE experiment at an average instantaneous luminosity of about $2 \times 10^{-25}$~${\rm cm}^{-2}{\rm s}^{-1}$ and a hadronic interaction rate of $80-150$~${\rm Hz}$.
In total, the \xexe data sample consists of about $1.1 \times 10^6$ minimum bias (MB) events passing the event selection described below.
The MB interaction trigger is provided by two arrays of forward scintillators, named V0 detectors, with a pseudorapidity coverage of $2.8 < \eta < 5.1$ (V0A) and $- 3.7 < \eta < -1.7$ (V0C)~\cite{Abbas:2013taa}.
The V0 signal is proportional to the charged-particle multiplicity and is used to divide the \xexe sample in centrality classes defined in percentiles of the hadronic cross section~\cite{Abelev:2013qoq, ALICE-PUBLIC-2018-003, Acharya:2018hhy}.
The analysis is carried out in the centrality classes \centint{0}{5}, \centint{5}{10}, \centint{10}{20}, \centint{20}{30}, \centint{30}{40}, \centint{40}{50}, \centint{50}{60}, \centint{60}{70}, \centint{70}{90}.
In order to reduce the statistical uncertainty, the \pphi measurements are obtained in wider centrality classes \centint{0}{10}, \centint{10}{20}, \centint{20}{30}, \centint{30}{40}, \centint{40}{50}, \centint{50}{70}, \centint{70}{90}.
The most central (peripheral) collisions are considered in the \centint{0}{5} (\centint{70}{90}) class.
The \centint{90}{100} centrality bin is not included in the analysis since it is affected by the contamination of electromagnetic processes ($\approx$~20\%).
In addition, as described in~\cite{PhysRevC.101.044907, Acharya:2018eaq}, an offline selection of the events is applied to remove the beam-background events.
It combines the V0 timing information and the correlation between the sum and the difference of times measured in each of the Zero Degree Calorimeters (ZDCs) positioned at $\pm$~112.5~m from the interaction point~\cite{Abelev:2014ffa}.
Due to the low instantaneous luminosity (with an average collision probability per bunch crossing of $\mu~\approx~0.0005$), the probability of having more than two events per collision trigger was sufficiently low that the so-called event pileup is considered negligible.

The central barrel detectors are located inside a solenoidal magnet providing a maximum magnetic field (B) of 0.5~T.
A magnetic field of 0.2~T can be set when operating the magnet in its low B field configuration.
The central barrel detectors are used to reconstruct tracks and measure their momenta, as well as to perform particle identification (PID).
Those exploited in this analysis are (from the interaction point outwards) the Inner Tracking System (ITS)~\cite{Aamodt:2008zz}, the Time Projection Chamber (TPC)~\cite{Alme:2010ke} and the Time Of Flight (TOF) detector~\cite{Akindinov:2013tea}.
With respect to previous analyses~\cite{PhysRevC.101.044907}, the low amount of collected data makes it impracticable to perform PID with the High Momentum Particle IDentification detector (HMPID)~\cite{Beole:1998yq}.

\begin{figure}[h]
  \centering
  \includegraphics[width=0.45\textwidth]{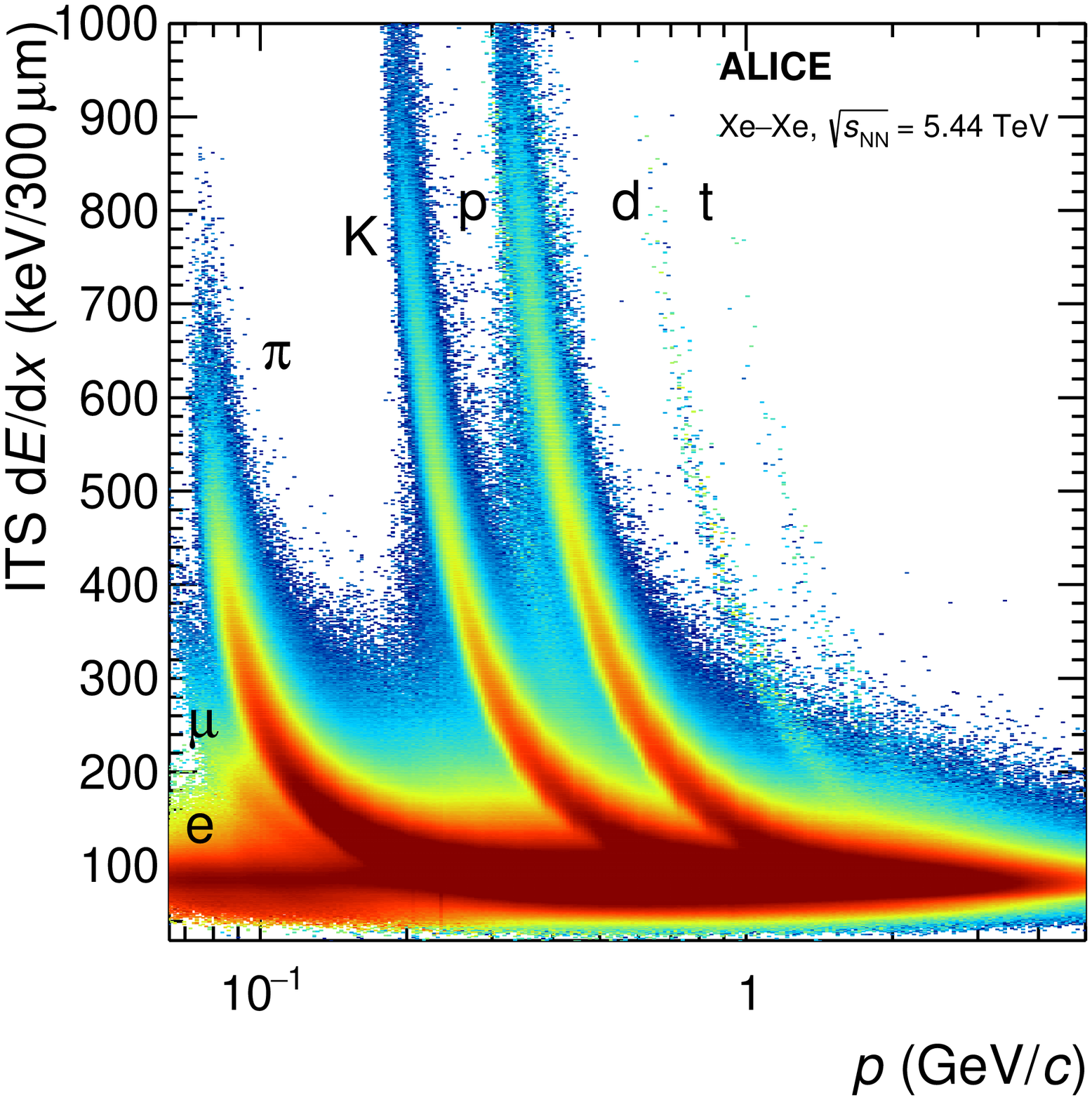}
  \includegraphics[width=0.45\textwidth]{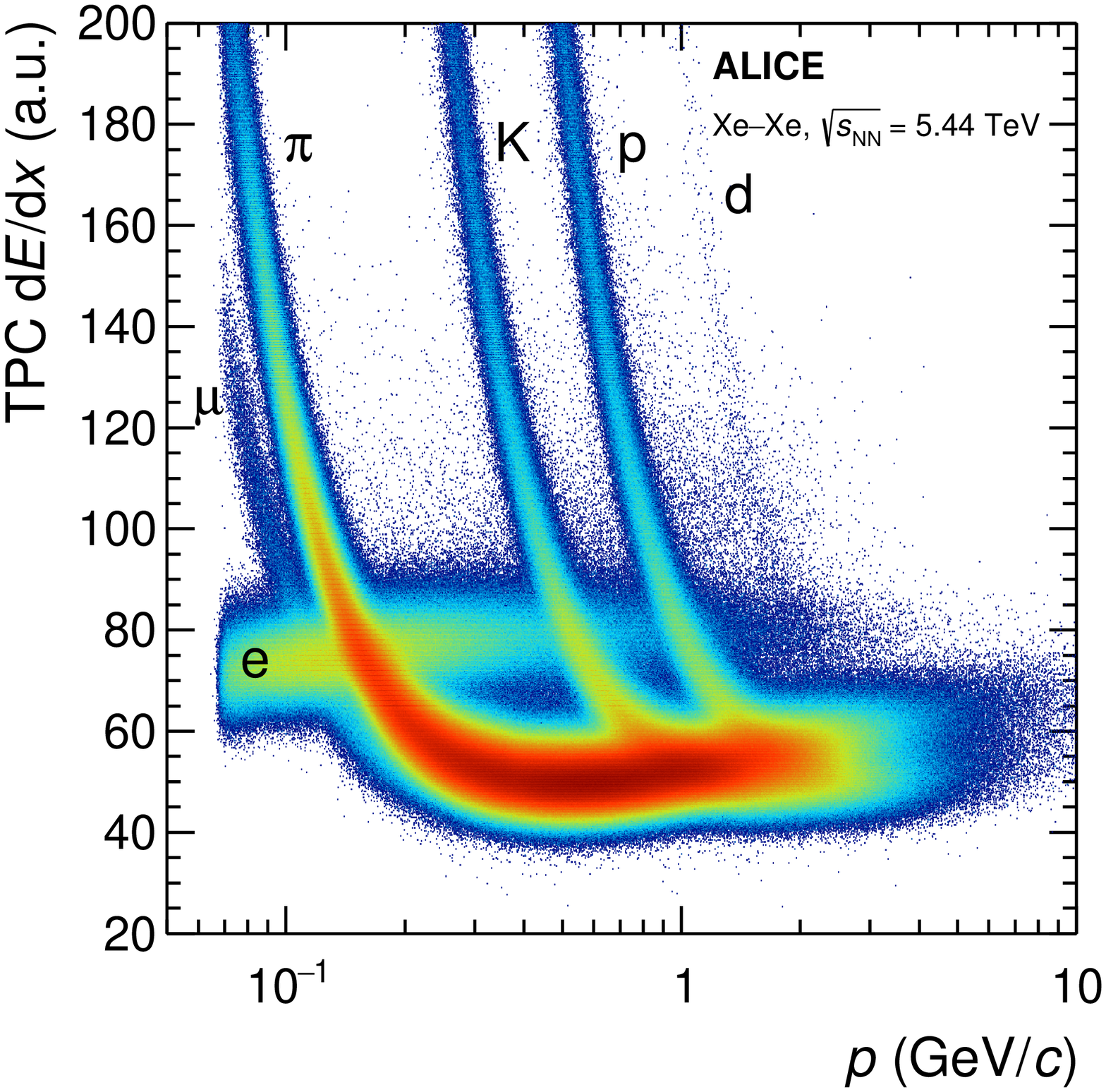}
  \caption{
    Distribution of the \dedx measured in the ITS (left) and TPC (right) detectors as a function of the reconstructed track momentum in \xexe collisions at \snnXeXe.
    The bands corresponding to the signals of \ppipm, \pkapm, \pprp and \pprm are well separated in the relevant momentum ranges.
    The good separation power obtained at low momentum is one of the key features for the measurements reported in this article.
  }
  \label{fig:itstpcperformance}
\end{figure}

The ITS is equipped with six layers of silicon detectors made of three different technologies: Silicon Pixel Detectors (SPD, first two layers from the interaction point), Silicon Drift Detectors (SDD, two middle layers) and Silicon Strip Detectors (SSD, two outermost layers).
It allows the reconstruction of the collision vertex, the reconstruction of tracks and the identification of particles at low momentum (\mom~$<$~1~\gevc) via the measurement of their specific energy loss (\dedx).
An ITS-only analysis can be performed by using a dedicated algorithm to treat the ITS as a standalone tracker, enabling the reconstruction and identification of low-momentum particles that do not reach the TPC.
The TPC, a cylindrical gas detector equipped with Multi-Wire Proportional Chambers (MWPC), constitutes the main central-barrel tracking detector and is also used for PID through the \dedx measurements in the gas.
The \dedx measurements obtained with the ITS and TPC detectors are shown in \Fig{fig:itstpcperformance}.
The \tofmeas measured with the TOF, a large area cylindrical detector based on Multigap Resistive Plate Chamber (MRPC) technology, combined with the momentum information measured in the TPC, is employed to identify particles at low and intermediate momenta ($\lesssim 5$ \gevc).

The events analysed in this article are chosen according to the selection criteria described in~\cite{PhysRevC.101.044907}.
The primary vertex is determined from tracks, including the track segments reconstructed in the SPD.
The position along the beam axis ($z$) of the vertex reconstructed with the SPD segments and of the one reconstructed from tracks are required to be compatible within 0.5 cm with a resolution of the SPD one better than 0.25 cm.
The position of the primary vertex along $z$ is required to be within 10 cm from the nominal interaction point.
These criteria ensure a uniform acceptance in the pseudorapidity region $|\eta| < 0.8$.

The results presented in this work refer to primary particles, defined as particles with a mean proper lifetime of $\tau > 1$~\cmc that are either produced directly in the interaction or from decays of particles with $\tau < 1$~\cmc, restricted to decay chains leading to the interaction point~\cite{ALICE-PUBLIC-2017-005}.
To reduce the contamination from secondary particles from weak decays and interactions in the detector material, as well as tracks with wrongly associated hits, similar selection criteria as described in~\cite{PhysRevC.101.044907, Acharya:2018eaq} are used and are summarised below.
Tracks reconstructed with both the TPC and the ITS are required to cross at least 70 TPC readout rows out of a maximum of 159 with a \chitwo normalised to the number of TPC space points (``clusters''), \chitwoCluster, lower than 4.
The ratio between the number of clusters and the number of crossed rows in the TPC has to be larger than 0.8.
An additional cut on the track geometrical length in the TPC fiducial volume is used as in~\cite{Acharya:2018eaq}.
Tracks are also required to have at least two hits in the ITS detector out of which at least one has to be in the SPD.
In addition, for the ITS-only analysis, the tracks must have at least three hits in the SDD + SSD layers.
The \chitwoCluster is also recalculated constraining the track to pass by the primary vertex and it is required to be lower than 36.
The same selection is also applied on the ITS points of the track: $\chitwoITS/\ndfITS < 36$.
For the ITS-only analysis, this selection is restricted to $\chitwoITS/\ndfITS < 2.5$.
Finally, the tracks are required to have a distance of closest approach (\dca) to the primary vertex along the beam axis lower than 2 cm.
A \pt-dependent selection is then applied to the \dca in the transverse plane (\dcaXY): $|\dcaXY| < 7 \sigmadcaXY$ where \sigmadcaXY is the resolution on the \dcaXY in each \pt interval.
Furthermore, the tracks associated with decay products of weakly decaying kaons (``kinks'') are rejected.
This selection is not applied for kaons studied via their kink decay topology.
The track selection criteria for kaons and pions from kinks will be described in the next paragraph.

The \xexe data were collected by operating the detector in its low B field configuration (${\rm B} =0.2$~T).
The lower magnetic field increases the probability of low momentum particles to cross the full detector thus extending the overall acceptance and reach of the analyses to lower \pt.
This allowed for the measurement of pions down to 50~\mevc for the first time at the LHC with respect to past publications~\cite{Abelev:2013vea,PhysRevC.101.044907} where the lowest \pt reach was to 100~\mevc.
While increasing the particle detection efficiencies at low momenta with respect to the standard field of 0.5~T, this configuration leads to a \pt resolution for ITS-only tracks that is worse by almost a factor 2 for \ppipm, \pkapm, \pprp and \pprm in their lowest \pt bin.
As a consequence, to achieve a reliable PID, an unfolding technique is used for ITS-only tracks to account for the resolution effects as it will be described in the next section.
On the contrary, the \tofmeas resolution and hence the performance of the TOF detector in terms of PID separation power is unaffected by the lower magnetic field.
Overall, the \tofmeas resolution is about 60~ps in central collisions.

\subsection{Pion, kaon and (anti-)proton analysis} \label{pikapr_analysis}
The particle identification for \ppipm, \pkapm, \pprp and \pprm relies on the signals measured in the ITS, TPC and TOF detectors.
This provides a separation between different particle hypotheses using track-by-track or statistical techniques.
In addition, \ppi and \pka are measured by reconstructing their weak decay (kink) topology~\cite{Abelev:2014ffa}.
Each of these identification techniques is best performing in a given \pt region, as reported in \Tab{tab:ptranges}, and all together cover a wide \pt interval of up to 5~\gevc.
The final spectra of each particle species are obtained by combining the single analyses.
The identification of \ppipm, \pkapm, \pprp and \pprm with ITS, TPC and TOF proceeds by evaluating the difference between the measured and expected signal (e.g. \dedx, \tofmeas) for a given species $i$ in terms of number-of-sigmas (\Nsigma):
\begin{equation}
  \Nsigma(i) = \frac{Signal_{\rm MEAS} - Signal_{\rm EXP}(i) }{\sigma(i)}
  \label{eq:nsgima}
\end{equation}
where $Signal_{\rm EXP}(i)$ is the expected signal and $\sigma(i)$ its expected standard deviation obtained under each particle mass hypothesis, as described in~\cite{Akindinov:2013tea, Abelev:2014ffa}.
A detailed description of such techniques and the measured separation power between the different particle species is shown for \pbpb collisions in~\cite{PhysRevC.101.044907} and it is unchanged for this data set.

\begin{table}[htbp]
  \centering
  \caption{
    Transverse momentum intervals and the corresponding PID methods for pions, kaons and (anti-)protons.
  }
  \begin{tabular}{lccccc} 
    \toprule
    Technique & \ppipm (\gevc) & \pkapm (\gevc) & \pprp and \pprm (\gevc) \\
    \midrule
    ITS       & 0.05$-$0.6     & 0.2$-$0.5      & 0.3$-$0.6               \\
    TPC       & 0.35$-$0.6     & 0.25$-$0.35    & 0.55$-$0.75             \\
    TOF       & 0.45$-$5.0     & 0.45$-$ 4.0    & 0.65$-$5.0              \\
    Kinks     & 0.3$-$0.95     & 0.3$-$5.0      & $-$                     \\
    \bottomrule
  \end{tabular}
  \label{tab:ptranges}
\end{table}
\paragraph{ITS analysis.}
The ITS can be used as a standalone low-\pt PID detector thanks to the particle energy loss (\dedx) measured in its four outermost layers~\cite{Abelevetal:2014dna}.
To correct for the detector resolution effects on the particle identification for \mom $\lesssim$ 1 \gevc, a Bayesian unfolding technique is employed with the RooUnfold package~\cite{DAgostini:1994fjx}.
The unfolding of the momentum distribution in \dedx slices (1.1 keV/300~$\mu$m each) is performed with a four-iteration procedure where the initial prior probability is taken from the generated momentum distribution in the Monte Carlo (MC) simulated events with \HIJING \cite{Wang:1991hta}.
A proper correction for detector inefficiencies and particle contamination is applied following the prescription in~\cite{DAgostini:1994fjx}.
The unfolded momentum (\momTRUE) corresponding to the maximum of the conditional probability P(\momTRUE $|$ \momMEAS) for a given measured momentum \momMEAS is considered for the evaluation of the expected signal in the \Nsigma approach (see \Eq{eq:nsgima}).
Based on this, the plane (\momTRUE; \dedx) is divided into identification regions where each point is assigned a unique particle identity.
The identity of a track is assigned based on the difference between the measured \dedx and the one computed under each mass hypothesis.
The hypothesis which gives the smallest distance is used, thereby removing the sensitivity to the parameterisation of the \dedx resolution.
A further selection $\abs{\Nsigma^{\rm \pi}} < 2$ rejects electrons in the pion identification.

To calculate the unfolded \pt distributions (vs \ptTRUE), the Bayesian unfolding is also applied to the raw \ptMEAS distributions of each species.
In this case, the initial prior probability for the unfolding is taken from the generated \pt distributions of each species in the MC and the number of iterations is kept to four so as to minimize the statistical fluctuations (different numbers are considered for the systematic uncertainty evaluation).

With this method it is possible to identify \ppipm, \pkapm, \pprp and \pprm in the following \pt ranges, respectively: \momint{0.05}{0.6}, \momint{0.2}{0.5} and \momint{0.3}{0.6}.
This also allows for the reduction of the contamination due to other particle species.
For the first time at the LHC, thanks to the low magnetic field configuration the \pt reach of the pion spectra is extended down to 50 \mevc with a contamination from electrons of about 30\%.
To this purpose, a detailed study in the low momentum region was carried out in different rapidity intervals to verify the stability of the measurement (as it will be explained in section~\ref{systematics}).
\paragraph{TPC and TOF analyses.}
The identification with the TPC and TOF detectors mostly follows the procedure developed in~\cite{PhysRevC.101.044907} with some adaptations.
In both cases, the response of the PID signal was tuned for the lower magnetic field configuration.
The raw yield of particles is extracted in each \pt interval via a statistical unfolding.
In particular, for the TOF analysis templates obtained with a data-driven approach are used.
An additional template is used to take into account the signal component due to the TPC-TOF track mismatch.
The excellent PID performance achieved with both detectors allowed a continuous separation of pions from kaons and kaons from (anti-)protons in a wide interval of \pt as reported in \Tab{tab:ptranges}.
\paragraph{Kink analysis.}
Charged kaons and pions can also be identified by reconstructing their weak decay topology (kink topology) defined as secondary vertices with two tracks (mother and daughter) having the same charge.
The kink topology is analysed inside the TPC volume within a radius of 110--220 cm.
Details about the kaon identification algorithm based on the kink topology can be found in~\cite{Aamodt:2011zj, Abelev:2014ffa, PhysRevC.101.044907, Adam:2015qaa}.
In this article, the identification of pions via their kink decay topology is reported for the first time at the LHC.

The identification of kaons from kink topology and their separation from pion decays is based on the two-body decay kinematics.
The method allows for the extraction of kaon and pion spectra on a track-by-track basis.
Both particles decay into $\mu+\nu_{\mu}$ with branching ratios (\BR) of 63.55\% (\pka) and 99.99\% (\ppi)~\cite{Zyla:2020zbs}.
For this decay channel, the transverse momentum of the charged daughter particle with respect to the direction of the mother track (\qt), has an upper limit of 236~\mevc for kaons and 30~\mevc for pions.
Taking into account that the upper limit of \qt for the decay $\pkapm\rightarrow\pi^\pm+\pi^0$ (with $\BR = 20.66\%$~\cite{Zyla:2020zbs}) is 205~\mevc, an effective separation of kaons from pions can be achieved by selecting kinks with $\qt > 40$~\mevc.
Further selections are applied to reach a purity of kaons higher than 95\%:
($i$) \qt $>$ 120 \mevc in order to discard pion and 3-body kaon decays,
($ii$) a kink radius in the transverse plane between 110 and 205 cm,
($iii$) at least 20 TPC clusters for the mother track,
($iv$) a decay angle greater than \degrees{2} in order to remove fake kinks from particles that are wrongly reconstructed as two separate tracks, and
($v$) a kink decay angle, at a given mother momentum, between the maximum decay angle for pion to muon ($\mu +\nu_{\mu}$ decay)
and the maximum decay angle of kaon to muon ($\mu +\nu_{\mu}$ decay).
Finally, identified kaons from kinks are accepted if the mother track is found to have a \dedx within $3.5\sigma$ around the expected Bethe-Bloch value for kaons.

The charged pions that are identified via their kink decay topology show a purity higher than 97\%.
Similar selection criteria as for kaons are used except for $10 < \qt < 40$~\mevc (the most effective cut) and with the requirement of a decay angle smaller than the maximum decay angle of $\ppi\rightarrow\mu+\nu_{\mu}$.
The difference in the \qt selection for kaon and pion identification is due to their different decay angles to a muon at equal mother momentum.
The maximum decay angle of a kink mother track with momentum $\mom = 1.5$~\gevc is \degrees{2} for the pion to muon decay while \degrees{50} for the kaon to muon decay, because of the mass difference of the mother particles.
This feature restricts the pion identification below $\mom = 1.5$~\gevc.

\subsubsection{Corrections for efficiency and feed-down}
The \pt distributions of \ppipm, \pkapm, \pprp and \pprm are obtained by correcting the raw spectra for PID efficiency, misidentification probability, acceptance and tracking efficiencies as performed in~\cite{PhysRevC.101.044907} for the ITS, TPC, TOF and kink analyses.
The efficiencies are obtained from Monte Carlo simulated events generated with \HIJING.
The propagation of particles through the detector is simulated with the \GEANTT transport code~\cite{Brun:1119728} where the detector characteristics and data-taking conditions are precisely reproduced.
Thanks to the lower magnetic field of the \xexe data sample, a tracking efficiency of about 2\% (2.4\%) is reached at the lowest \pt point (\pt~=~50~\mevc) for pions in the most central (peripheral) bin compared to an efficiency lower than 1\textperthousand{} at full field.
It is known~\cite{Abelev:2013vea, Aamodt:2010dx, PhysRevC.101.044907} that the energy loss of low-\pt \pprm in the detector material and the cross section of low-\pt \pkam are not well reproduced in \GEANTT.
For this reason, a correction of the efficiency is estimated using \GEANTF~\cite{Agostinelli:2002hh} and \FLUKA~\cite{Battistoni:2007zzb}, respectively, in which these processes are reproduced more accurately.
The corrections amount to about 10\% and 4\% for \pprm and \pkam, respectively, in the lowest \pt bin (see \Tab{tab:ptranges}).
The PID efficiency and the misidentification probability are estimated in the simulation by requiring the simulated data to reproduce the real PID response for each detector included in this analysis.

The raw distributions are further corrected for the contribution of secondary particles (feed-down) mainly due to weak decays of \pkazero (affecting \ppipm), \plam and \psigp (affecting \pprp and \pprm).
Secondary protons coming from the detector material are also subtracted from the raw spectrum.
The estimation of this correction factor is data-driven since the event generators underestimate the strangeness production and, as already mentioned, the transport codes do not provide a precise description of the interaction of low-\pt particles with the detector material.
For each analysis, the reconstructed \dcaXY distributions for each particle species are fitted in each \pt interval with three contributions (as templates) extracted from the Monte Carlo simulation: primary particles, secondary particles from weak decays of strange hadrons and secondary particles produced in the interaction with the detector material, similarly to what is reported in~\cite{Abelev:2013vea, PhysRevC.101.044907}.
Finally, the spectra are normalized to the total number of events analysed in each centrality class.
The spectra in the extended \pt range are obtained by combining those obtained with the single identification techniques.
In the \pt intervals where more analyses overlap, the combination is carried out by performing an averaged mean using the single systematic uncertainties as weights.

\subsection{\texorpdfstring{\pphi}\ meson analysis} \label{phi_analysis}
The \pphi meson signal is reconstructed via invariant mass analysis by exploiting the decay channel into charged kaons, \pphi~$\rightarrow~\pkap\pkam$ (\BR~=~0.492~$\pm$~0.005~\cite{Zyla:2020zbs}).
The analysis follows a consolidated technique described extensively in~\cite{Adam:2016bpr, Adam:2017zbf, Acharya:2019qge}.
Candidate kaons are identified based on the variable defined by \Eq{eq:nsgima} for the \dedx sampled in the TPC (\NsigmaTPC) or the \tofmeas measured by the TOF (\NsigmaTOF).
More precisely, a track associated with a hit in the TOF detector is identified as a K if $\NsigmaTOFabs < 3$ and $\NsigmaTPCabs < 5$.
If a track does not reach the TOF detector and no \tofmeas measurement is available, only the information of the TPC is used by requiring that
$\NsigmaTPCabs < 2$ for $\pt > 0.4~\gevc$,
$\NsigmaTPCabs < 3$ for $0.3 < \pt < 0.4~\gevc$, and
$\NsigmaTPCabs < 5$ for $\pt < 0.3~\gevc$.
Within each event, identified kaons are combined in oppositely-charged pairs (``unlike-sign'') to extract the invariant mass ($\minv$) distribution of the signal.
To estimate the background from uncorrelated pairs, an event mixing technique is used, which consists in building the invariant mass distribution of $\pkap\pkam$ pairs from five different events with similar centrality (within 5$\%$) and a similar vertex position along the beam axis (within 1 cm).
Only same-event and mixed-event pairs with rapidity $\abs{y} < 0.5$ are selected.
The mixed-event background is normalised to the integral of the unlike-sign distribution in the invariant mass interval $1.07\leq\minv\leq1.1~\gevcsq$ and then subtracted.
The resulting distribution exhibits a clear peak centered at the nominal mass of \pphi~\cite{Zyla:2020zbs}, on top of a low residual background.
The \pphi signal peak is fitted with a Voigtian function (as in~\cite{Abelev:2014uua}), which is the convolution of a Breit--Wigner, describing the characteristic shape of the resonance state, and a Gaussian, taking into account the detector resolution.
The resonance width is fixed to the nominal value of $\Gamma_\pphi = 4.26~\mevcsq$~\cite{Zyla:2020zbs}, whereas the mass and the mass resolution $\sigma_\pphi$ are left as free fit parameters.
The mass resulting from the fit is consistent with the nominal value of the \pphi mass 
reported in~\cite{Zyla:2020zbs}.
The $\sigma_\pphi$ parameter ranges from $\approx$~1.5~\mevcsq at $\pt = 0.5-1~\gevc$ to $\approx$~2.5~\mevcsq at \pt~=~10~\gevc, and it is consistent with the mass resolution extracted from Monte Carlo simulations of the full detector setup and reconstruction chain.
The residual background is parameterised with a linear function.
The fit is performed in the range $0.994 < \minv < 1.07$ \gevcsq.
This procedure is repeated for each \pt and centrality interval.

The \pt-differential yields obtained with the described procedure are corrected for efficiency and acceptance, as described in~\cite{Acharya:2019qge}.
The corrections are obtained from a Monte Carlo simulation where events are generated with \HIJING~\cite{Wang:1991hta} and particles are transported through a detailed simulation of the \ALICE detector with the \GEANTT transport code~\cite{Brun:1119728}.
The selection criteria for \pphi candidates are the same in Monte Carlo and data.

\subsection{Systematic uncertainties} \label{systematics}
The calculation of the systematic uncertainties follows the procedure performed already for previous analyses~\cite{PhysRevC.101.044907, Abelev:2013vea, Adam:2015qaa, Abelev:2014uua, Adam:2017zbf}.
The main sources of systematic uncertainties for each particle species are summarised in \Tab{tab:sys_pikapr} (\ppipm, \pkapm, \pprp and \pprm) and in \Tab{tab:sys_phi} (\pphi).


\newif\ifExtend
\Extendtrue
\Extendfalse

\begin{table}[htbp]
  \centering
  \caption{
    Main sources and values of the relative systematic uncertainties (expressed in \%) of the \pt-differential yields of \ppipm, \pkapm, \pprp and \pprm obtained in the analysis of \xexe collisions.
    The first section is common to all the analyses, the analysis specific uncertainties are listed separately.
    When two values are reported, they correspond to the lowest and highest \pt bin respectively, considering the maximum contribution among the various centrality classes.
    If only one value is reported, the systematic uncertainty is not \pt-dependent.
    For certain sources, the centrality is specified when a larger dependence on centrality is observed.
    The maximum total systematic uncertainties (among all centrality classes) are shown.
    The total uncertainty refers to the uncertainty of the combined results (see text).
  }
  \ifExtend
    \resizebox{\columnwidth}{!}{%
      \fi
      \begin{tabular}{lcccc} 
        \toprule
        Effect                                             & \ppipm (\%) & \pkapm (\%) & \pprp and \pprm (\%) \\
        \midrule
        ITS$-$TPC matching efficiency (\centint{0}{5})     & 2.2$-$0.4   & 2.2$-$0.4   & 2.2$-$0.4            \\
        ITS$-$TPC matching efficiency (\centint{40}{50})   & 3.0$-$1.2   & 3.0$-$1.2   & 3.0$-$1.2            \\
        ITS$-$TPC matching efficiency (\centint{70}{90})   & 2.8$-$0.6   & 2.8$-$0.6   & 2.8$-$0.6            \\
        Material budget                                    & 1.6$-$0.2   & 1.3$-$0.4   & 2.9$-$0.1            \\
        Hadronic interaction cross section                 & 2.5$-$2.4   & 2.7$-$1.8   & 4.6                  \\
        \midrule
        \multicolumn{4}{c}{ITS analysis}                                                                      \\
        PID                                                & 1.4$-$3.1   & 1.4$-$7.7   & 1.2$-$0.7            \\
        Track selection                                    & 4.7$-$4.4   & 6.0$-$6.7   & 9.8$-$7.9            \\
        \EcrossB                                           & 3.0         & 3.0         & 3.0                  \\
        Unfolding iterations                               & 5.5$-$2.2   & 6.1$-$5.2   & 13.7$-$2.3           \\
        Rapidity selection                                 & 7.0$-$3.0   & 3.0         & 10.0                 \\
        Feed-down correction                               & 3.2$-$3.2   & 3.0$-$3.0   & 3.0$-$3.0            \\
        Matching efficiency (\centint{0}{5})               & 1.2         & 1.2         & 1.2                  \\
        Matching efficiency (\centint{40}{50})             & 0.5         & 0.5         & 0.5                  \\
        Matching efficiency (\centint{70}{90})             & 2.0         & 2.0         & 2.0                  \\
        Hadronic interaction cross section (ITS tracks)    & 3.0$-$0.3   & 2.7$-$1.5   & 13.3$-$5.6           \\
        \midrule
        \multicolumn{4}{c}{TPC analysis}                                                                      \\
        PID (\centint{0}{5})                               & 14.$-$14.4  & 3.3$-$15.0  & 4.3$-$19.5           \\
        PID (\centint{40}{50})                             & 5.4$-$5.3   & 2.0$-$7.4   & 0.8$-$9.5            \\
        PID (\centint{70}{90})                             & 3.9$-$4.6   & 2.1$-$6.6   & 1.0$-$4.8            \\
        Track selection                                    & 0.4$-$1.5   & 5.0$-$6.0   & 3.8$-$3.0            \\
        Feed-down correction                               & 0.5         & $-$         & 0.8$-$9.7            \\
        \midrule
        \multicolumn{4}{c}{TOF analysis}                                                                      \\
        PID                                                & 3.0$-$12.0  & 3.0$-$18.0  & 2.0$-$20.0           \\
        Track selection                                    & 1.5         & 1.5         & 1.8                  \\
        Matching efficiency                                & 1.2$-$5     & 4.5$-$5.0   & 5.3$-$5.0            \\
        Feed-down correction                               & 0.5         & $-$         & 9.7$-$0.4            \\
        \midrule
        \multicolumn{4}{c}{Kink analysis}                                                                     \\
        PID + reconstruction efficiency (\centint{0}{5})   & 2.6         & 1.7$-$6.0   & $-$                  \\
        PID + reconstruction efficiency (\centint{40}{50}) & 2.6         & 1.0$-$4.4   & $-$                  \\
        PID + reconstruction efficiency (\centint{70}{90}) & 1.6         & 2.7$-$4.7   & $-$                  \\
        Contamination (\centint{0}{5})                     & 1.0$-$4.0   & 0.5$-$5.3   & $-$                  \\
        Contamination (\centint{40}{50})                   & 1.0$-$2.0   & 0.5$-$3.2   & $-$                  \\
        Contamination (\centint{80}{90})                   & 1.0$-$2.0   & 0.5$-$3.0   & $-$                  \\
        \midrule
        Total                                              & 11.1$-$21.9 & 9.0$-$10.0  & 22.4$-$10.5          \\
        \bottomrule
      \end{tabular}
      \ifExtend
    }
  \fi
  \label{tab:sys_pikapr}
\end{table}

The main sources of systematic uncertainty affecting this analysis are: PID, feed-down correction, the imperfect description of the material budget in the Monte Carlo simulation, the knowledge of the hadronic interaction cross section in the detector material~\cite{PhysRevC.101.044907}, the ITS-TPC~\cite{Acharya:2018eaq} (accounted twice for the decay daughters of the \pphi) and TPC-TOF matching efficiencies, the track selection, the unfolding iterations and the rapidity selection for the ITS.
The uncertainties for track selection refer to the quality requirements based on the number of crossed rows in the TPC, the number of clusters in the ITS, the \dcaXY and \dcaZ, and the \chitwoNDF of the reconstructed tracks.
To estimate these uncertainties, a variation of the standard selection criteria is performed and the ratio between the corrected spectra with modified selection criteria and the ones with standard requirements is calculated, as performed in~\cite{PhysRevC.101.044907}.
For the uncertainty related to the number of iterations in the Bayesian unfolding for the ITS analysis, a similar approach is followed where the number of iterations is changed from 4 (default) to 3, 5, 7 and 9.
The uncertainties related to PID are evaluated by comparing different techniques (e.g. statistical unfolding versus track-by-track \Nsigma selection).
In addition, for the \pphi, a detailed study of the yield extraction procedure was carried out by investigating the effect of variations in the signal shape parameters, the background shape and the fit range, as performed in~\cite{Abelev:2014uua}.
The uncertainties of the detector material budget are estimated by changing the material budget in the simulation with the \GEANTT transport code by $\pm$7\% as in~\cite{PhysRevC.101.044907, Abelev:2012cn}.
The uncertainty of the hadronic interaction cross section is calculated by comparing the efficiencies in different transport codes (\GEANTT, \GEANTF, \FLUKA) following the prescription given in~\cite{Abbas:2013rua}.
Finally, the uncertainties on the feed-down are determined by varying the range of the template fit to the \dcaXY distributions.

For the ITS analysis, a systematic uncertainty is introduced to take into account the shift of the cluster positions caused by the Lorentz force (\EcrossB effect), as described in~\cite{PhysRevC.101.044907}.
For the kink analysis, the systematic uncertainties are estimated by comparing the standard spectra with the ones obtained by varying the selection criteria on the decay product transverse momentum, the minimum number of TPC clusters and the kink radius.

Finally, the systematic uncertainties on the very low \pt region of the spectra are higher compared to previous analyses~\cite{PhysRevC.101.044907, Abelev:2013vea} because of the lower momentum resolution in the reduced magnetic field.
Nonetheless, the uncertainty on the pion measurement below 100 \mevc is below 12\%.
In addition, the limited statistics of the \xexe data sample restricts the detectors and techniques that can contribute to the PID at higher momenta, excluding the HMPID detector and the TPC energy loss measurement in the relativistic rise region.
This yields overall larger uncertainties with respect to previous \ALICE measurements in other collision systems.
At 3~\gevc the uncertainties are approximately twice as large with respect to~\cite{PhysRevC.101.044907} for \ppipm, \pkapm, \pprp and \pprm.
\vspace{0.5cm}

\begin{table}[htbp]
  \centering
  \caption{
    Main sources and values of the relative systematic uncertainties (expressed in \%) of the \pt-differential yields of \pphi obtained in the analysis of \xexe collisions.
    When two values are reported, they correspond to the lowest and highest \pt bin respectively, considering the maximum contribution among the various centrality classes.
    If only one value is reported, the systematic uncertainty is not \pt-dependent.
    The maximum total systematic uncertainties (among all centrality classes) are shown.
  }
  {%
    \begin{tabular}{lcccc} 
      \toprule
      Effect                        & \pphi (\%)          \\
      \midrule
      \BR                           & 1                   \\
      ITS$-$TPC matching efficiency & \interval{6.4}{11}  \\
      Track cuts                    & \interval{2.2 }{ 4} \\
      PID                           & \interval{2 }{ 12}  \\
      Hadronic interaction          & \interval{2.2}{0}   \\
      Material budget               & \interval{1.0}{0}   \\
      Yield extraction              & \interval{5}{15}    \\
      \midrule
      Total                         & \interval{10}{20}   \\
      \bottomrule
    \end{tabular}
  }
  \label{tab:sys_phi}
\end{table}

\section{Results and discussion} \label{sec:results}
\label{Sec.:Results}

\subsection{Transverse momentum spectra}

\begin{figure}[h]
  \centering
  \includegraphics[width=0.9\textwidth]{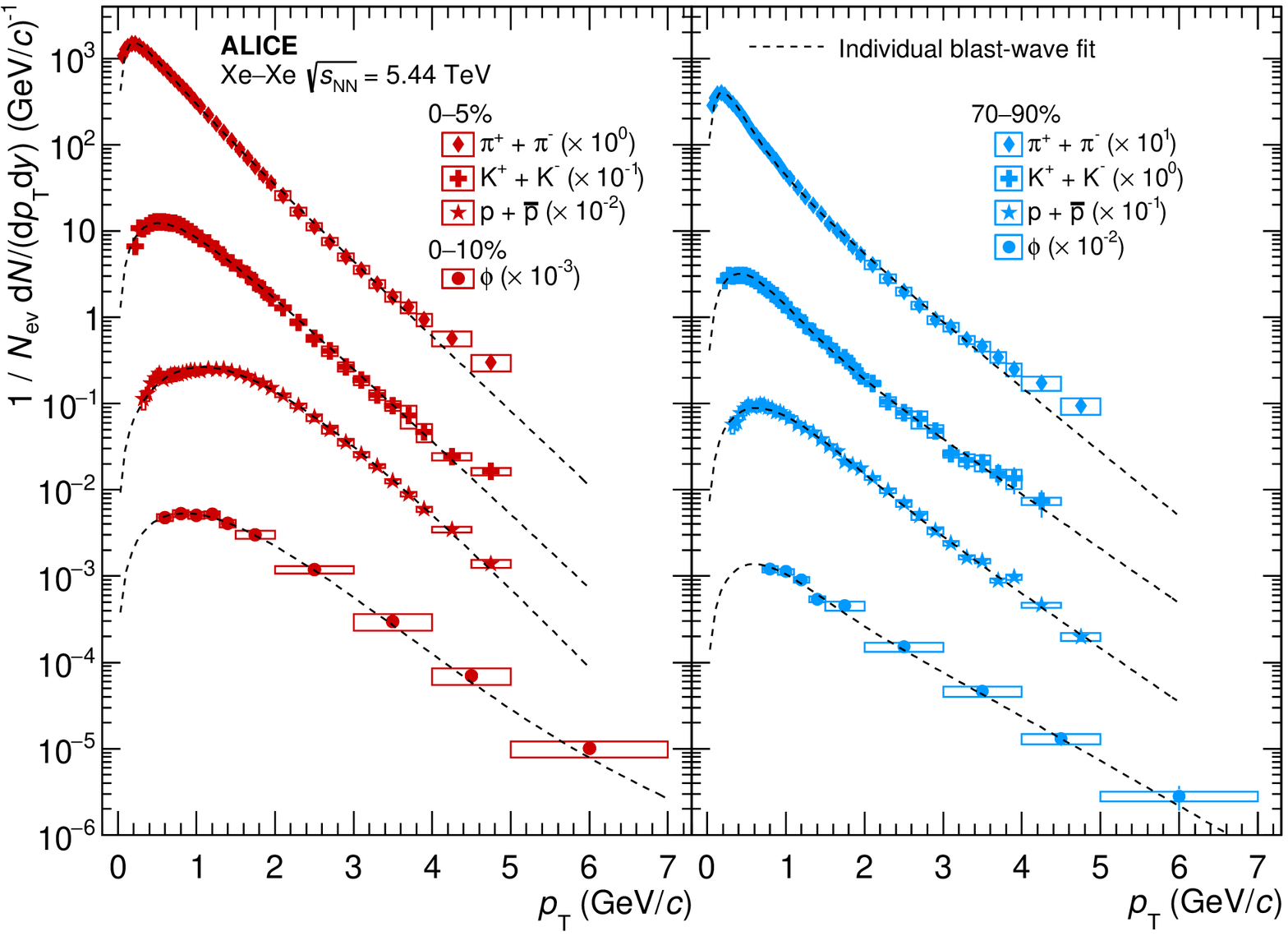}
  \caption{
    \pt distributions of \ppipm, \pkapm, \pprp, \pprm, \pphi as measured in central (left) and peripheral (right) \xexe collisions at \snnXeXe.
    The statistical and systematic uncertainties are shown as error bars and boxes around the data points.
  }
  \label{fig:spectra}
\end{figure}

The \ppipm, \pkapm, \pprp, \pprm and \pphi \pt spectra obtained after all corrections are shown for central and peripheral collisions in \Fig{fig:spectra}.
Each spectrum is individually fitted with a Blast-wave function~\cite{Schnedermann:1993ws}, shown with dashed lines.
The integrated yield \avdndy and the mean transverse momenta \meanpt are calculated from the measured spectra and the extrapolation of the Blast-wave functions in the unmeasured regions.
As performed in previous analyses~\cite{Abelev:2013vea, PhysRevC.101.044907}, the systematic uncertainties for both \avdndy and \meanpt are evaluated by shifting the data points up and down within their systematic uncertainty to obtain the softest and hardest spectra.
An additional contribution is given by the extrapolation to \pt~=~0~\gevc where different functions (\mt-exponential, Fermi-Dirac, Bose-Einstein, Boltzmann) were used for the calculation.
The uncertainty on the extrapolation for the most central collisions is found to be $\sim 1$\% for pions and kaons, $\sim 5$\% for protons and $\sim 2$\% for \pphi.

\begin{figure}
  \centering
  \includegraphics[width=.6\textwidth]{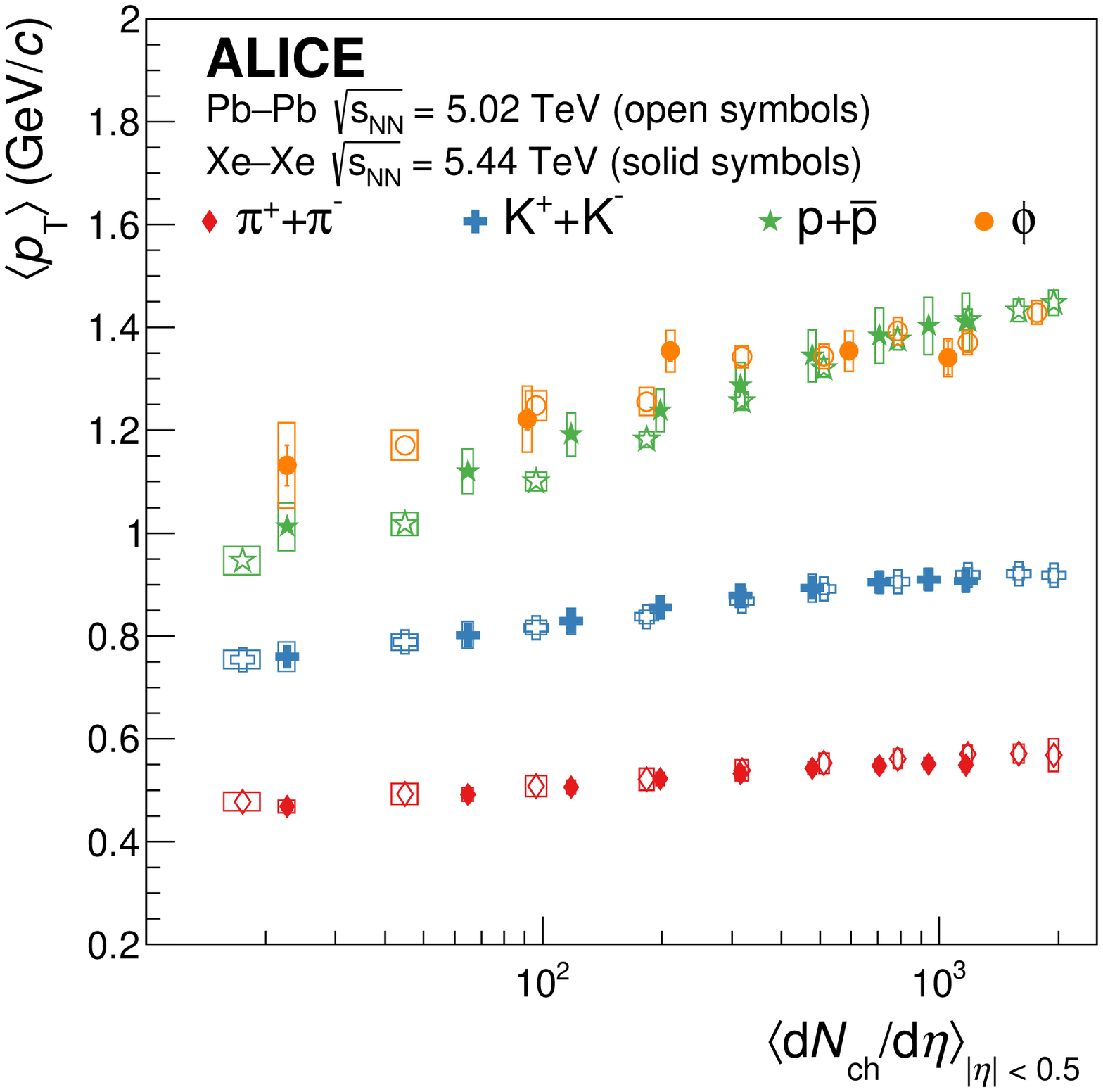}
  \caption{
    Mean \pt of pions, kaons, (anti-)protons and \pphi as a function of the charged-particle multiplicity density in \xexe collisions at \snnXeXe and \pbpb collisions at \snnF~\cite{PhysRevC.101.044907, Acharya:2019qge}.
    The statistical and systematic uncertainties are shown as error bars and boxes around the data points.
  }
  \label{fig:MeanPt}
\end{figure}

As already observed in \pbpb and also in small collision systems~\cite{PhysRevC.101.044907,Acharya:2018orn,Abelev:2013haa}, the \meanpt rises with increasing centrality and multiplicity (\avdNchdeta).
This hardening is significantly more pronounced for heavier particles.
For instance, the maximum of the \ppr spectrum shifts from $\pt\approx 0.8$~\gevc in peripheral to $\pt\approx 1.4$~\gevc in central collisions, while for pions the shift is much smaller.
This feature is generally considered as a consequence of the radial expansion of the system.
The comparison of \meanpt as a function of charged-particle multiplicity for \pbpb and \xexe collisions, shown in \Fig{fig:MeanPt}, clearly demonstrates that this effect is entirely driven by the multiplicity and not by the collision geometry.
Most notably, the \meanpt values of protons and \pphi differ in peripheral (low \dNchdeta) \xexe and \pbpb collisions, but reach similar values in semi-central and central collisions.
This behaviour is expected due to the small mass difference of these two particles if the spectral shape is more and more dominated by radial flow with increasing centrality.

\begin{figure}
  \centering
  \includegraphics[width=0.9\textwidth]{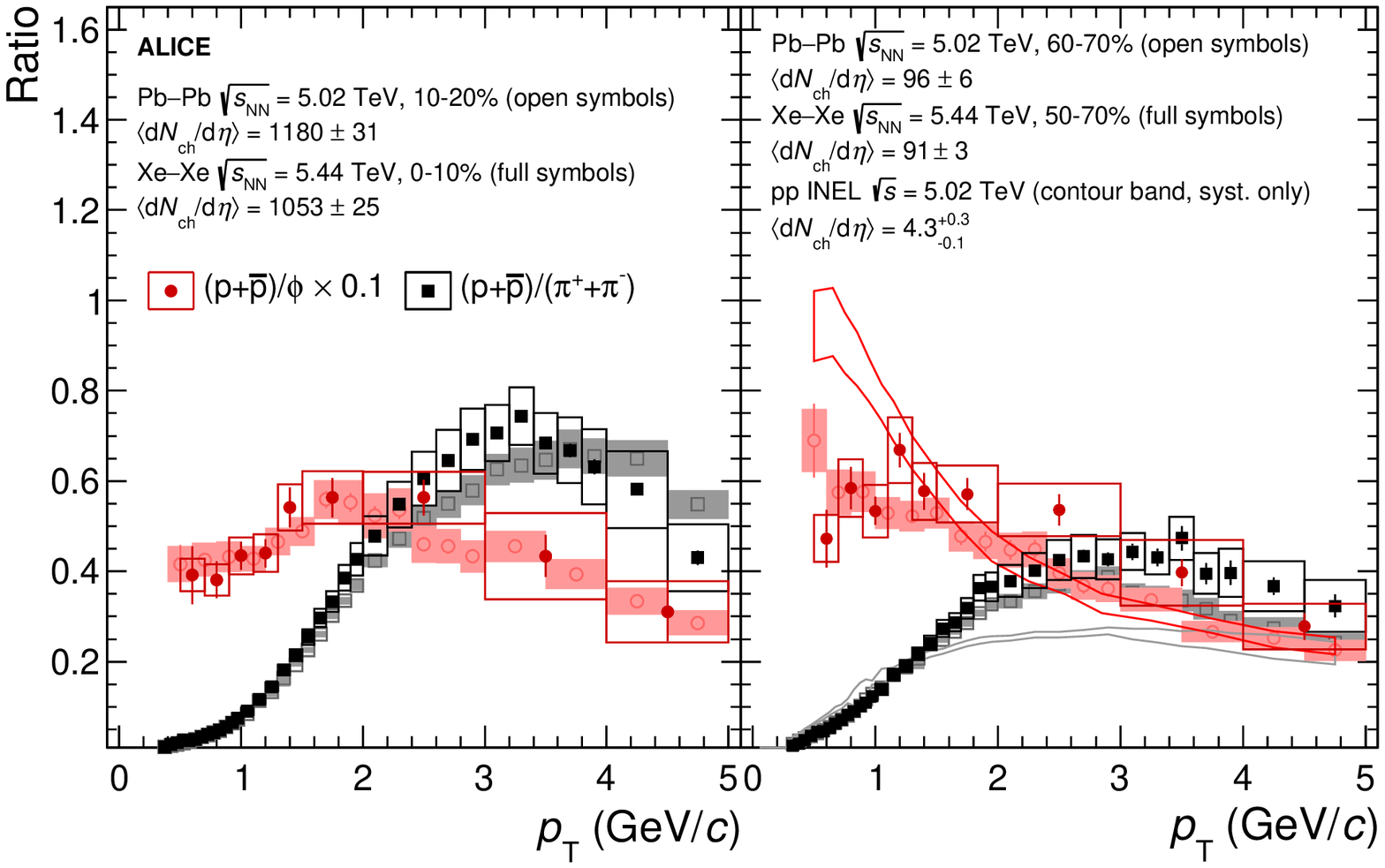}
  \caption{
    Left: proton-to-phi and proton-to-pion \pt-differential ratios in \centint{0}{10} central \xexe collisions at \snnXeXe and \centint{10}{20} central \pbpb collisions at \snnF~\cite{PhysRevC.101.044907}.
    Right: proton-to-phi and proton-to-pion \pt-differential ratios in \centint{50}{70} \xexe collisions at \snnXeXe and \centint{60}{70} \pbpb collisions at \snnF~\cite{PhysRevC.101.044907, Acharya:2019qge}.
    The two selected groups of centrality bins have similar \avdNchdeta (see text for details).
    The statistical and systematic uncertainties are shown as error bars and boxes around the data points.
    The \pt-differential ratios measured in \pp collisions at \sF~\cite{PhysRevC.101.044907, Acharya:2019qge} are also shown in the right panel for comparison.
    The bands represent the systematic uncertainties alone.
  }
  \label{fig:ptratios}
\end{figure}

The mass-dependent radial flow naturally explains in central collisions the so-called baryon-to-meson enhancement at low to intermediate \pt ($\lesssim 5$~\gevc) observed in the light-flavour sector~\cite{PhysRevC.101.044907}.
This effect is seen in \Fig{fig:ptratios} where the \rppi ratio shows a maximum at around \momint{3}{4}.
Considering the most central \xexe collisions, which have a multiplicity similar to \centint{10}{20} \pbpb collisions at \snnF~\cite{PhysRevC.101.044907}, the \rppi ratio at the peak is enhanced by a factor of about 3 with respect to \pp collisions at the same energy.
Instead, in peripheral \pbpb collisions the effect of the radial flow is less evident and a \pt-dependence similar to the one found in \pp is observed.
Therefore, the measurements shown in \Fig{fig:ptratios} for peripheral collisions suggest that this consideration might hold true also in \xexe collisions.
Another explanation for the baryon-to-meson enhancement advocates quark recombination~\cite{Greco:2003xt, Fries:2003vb} as the dominant production mechanism for baryons at intermediate momenta.
In this picture, the production of baryons is enhanced at intermediate momenta as it is more likely to combine three soft quarks (with $p_{\rm T, q} = {\pt}/3$) into a baryon in order to reach a given momentum \pt than to produce a meson via quark-antiquark pair (each with $p_{\rm T, q} = {\pt}/2$).
However, the \rpphi ratio displayed in \Fig{fig:ptratios} is rather independent of \pt as expected in the radial flow picture.
Although their quark content is different, p and \pphi have similar masses, indicating that this is the main variable in the determination of the spectral shape.
Nevertheless, as discussed in~\cite{Minissale:2015zwa}, the same model including radial flow and coalescence plus fragmentation is able to describe both \rppi and \rpphi in central \pbpb collisions showing that both radial flow and recombination play a role.

\begin{figure}
  \centering
  \includegraphics[width=0.6\textwidth]{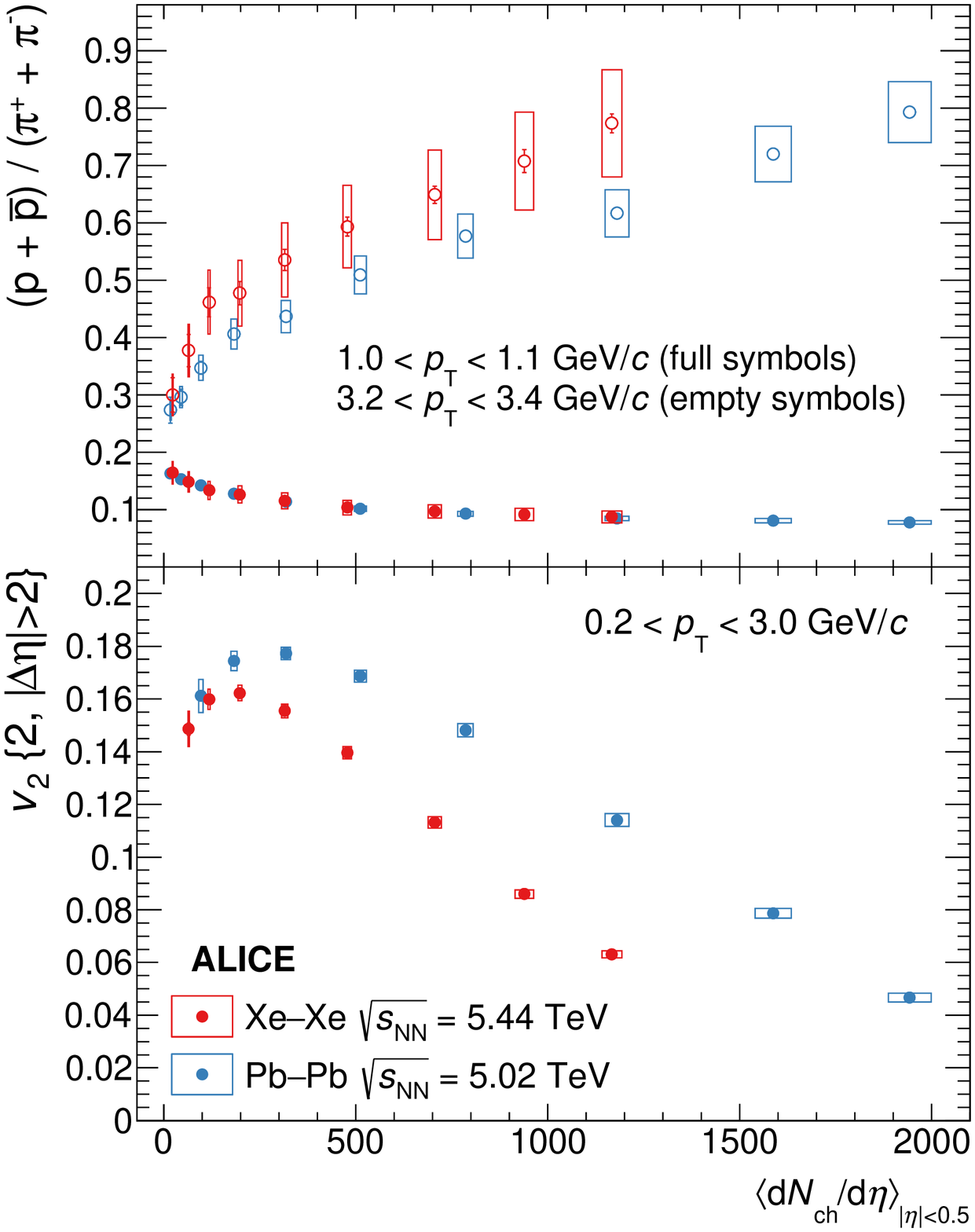}
  \caption{
    Proton-to-pion ratio as a function of charged particle multiplicity density in two \pt intervals for \xexe and \pbpb collisions at \snn~=~5.44 and 5.02 TeV.
    In the bottom panel, the flow coefficient \vtwotwo is plotted for the same collision systems~\cite{Acharya:2018zuq, Acharya:2018ihu} as a function of charged particle multiplicity density.
    The statistical and systematic uncertainties are shown as error bars and boxes around the data points.
  }
  \label{fig:ptopimulti}
\end{figure}

A direct comparison of the \xexe with \pbpb collisions allows the study of systems with the same charged particle density and different initial eccentricity: semi central \pbpb collisions have the same multiplicity as central \xexe collisions, however, the initial eccentricity is smaller in the latter case.
A difference in the initial eccentricity affects the hydrodynamic expansion, eventually leading to a different elliptic flow of the charged particles.
This is best illustrated in \Fig{fig:ptopimulti} which compares the elliptic flow coefficient \vtwotwo of charged particles (for details on the definition, see~\cite{Acharya:2018zuq, Acharya:2018ihu}) with the \rppi ratio.
Due to the large mass difference between protons and pions this ratio is very sensitive to radial flow effects.
Consequently, a depletion of this ratio at low transverse momenta and an enhancement at intermediate transverse momenta with increasing particle density is observed.
The magnitude of this effect is not only qualitatively, but also quantitatively, within uncertainties the same in \xexe and \pbpb collisions for similar charged particle densities.
In contrast, the \vtwo coefficient shows large differences between the two collision systems at similar particle densities, because it is dominantly influenced by the initial eccentricity.

\subsection{Hadrochemistry}

\begin{figure}
  \centering
  \includegraphics[width=.6\textwidth]{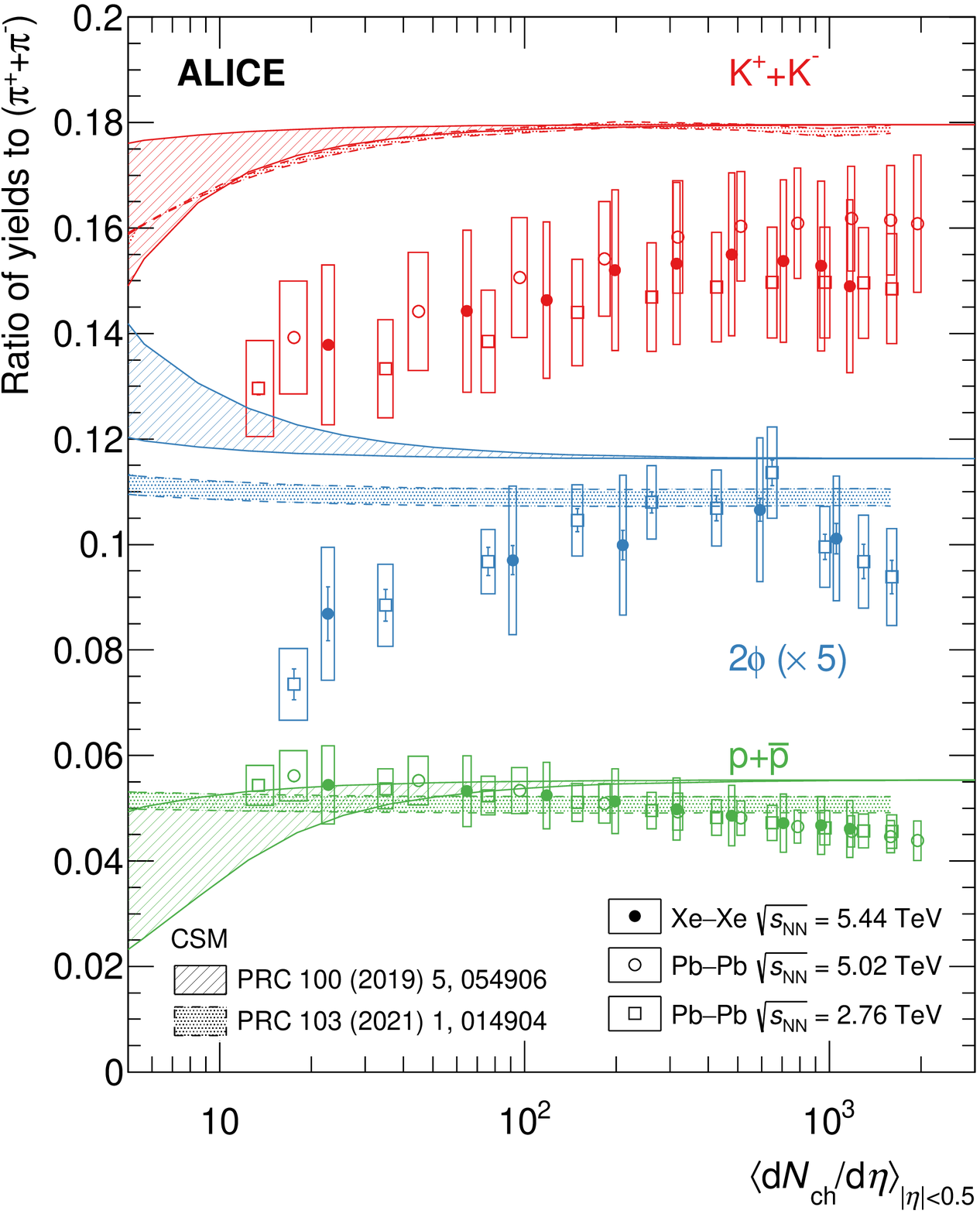}
  \caption{
    Ratio of kaon, proton and \pphi integrated yields to pion integrated yield as a function of the charged-particle multiplicity density for \xexe collisions at \snnXeXe and \pbpb collisions at \snnT~\cite{Abelev:2013vea, Abelev:2014uua} and 5.02~TeV~\cite{PhysRevC.101.044907, Acharya:2019qge}.
    The statistical and systematic uncertainties are shown as error bars and boxes around the data points.
    Predictions from the canonical statistical model (CSM) are shown as bands considering different correlation volumes~\cite{Vovchenko:privatecomm} (based on \cite{Vovchenko:2019kes}) and chemical freeze-out temperatures~\cite{Cleymans:2020fsc}.
    The correlation volume indicates the volume over which the strangeness conservation is imposed.
  }
  \label{fig:YieldHI}
\end{figure}

To investigate the particle chemistry, the \pt-integrated particle yields are determined in each centrality bin with the procedure described above for the \meanpt.
The resulting \avdndy values are summarised in \Tab{tab:dndyXeXe}.
The ratios of kaons, (anti-)protons, and \pphi to pions are shown in \Fig{fig:YieldHI} and compared with results from \pbpb collisions.
Similarly to the spectral shapes, also the particle yield ratios are comparable between \xexe and \pbpb collisions at similar charged-particle multiplicities.
The results reinforce two of the surprising features that were first observed in \pbpb collisions at the LHC energies and are now confirmed in a new heavy-ion collision system.
First, the \rppi-ratio values are around 0.05, significantly lower than those predicted before the LHC era~\cite{Abelev:2012wca}.
While the overall magnitude is understood as a consequence of the pion-nucleon phase-shift~\cite{Cleymans:2020fsc, Andronic:2018qqt} the decreasing trend with increasing centrality can be interpreted as a consequence of the antibaryon-baryon annihilation~\cite{Becattini:2014hla}.
The results presented in this article add constraints to the particle production mechanisms proposed to explain this observation.
The data reported in this work suggests that at LHC energies, particle production is not only independent of collision energy but also of the collision system when studied as a function of multiplicity.
Second, the \rphipi ratio shows an increasing trend from peripheral to central collisions with a hint of a decrease at higher multiplicities.
Notably, this increase appears to be slightly stronger for \rphipi with respect to \rkpi.
As shown in \Fig{fig:YieldHI}, this is not expected in canonical statistical hadronisation models~\cite{Vovchenko:2019kes, Cleymans:2020fsc}, which predict a constant or slightly decreasing trend since the net strangeness content $S$ of the \pphi is zero.
This feature is predicted from both models reported in \Fig{fig:YieldHI}, independent of the fact that the correlation volume over which the strangeness conservation is imposed is kept fixed in~\cite{Vovchenko:2019kes} and has a multiplicity dependence in~\cite{Cleymans:2020fsc}.
Future studies including the measurement of double-strange ($S = 2$) $\Xi$ baryons in \xexe collisions can determine across all available collision systems whether the increase for the \pphi is closer to $S = 1$ (such as kaons or lambdas) or $S = 2$ particles ($\Xi$).
The measurements of \pphi production in \pbpb collisions~\cite{Acharya:2019bli} indicate that the increase lies in between these two extremes.


\begin{table}
  {
    \begin{center}
      \caption{
        \avdndy of pions, kaons, (anti-)protons and \pphi for different centrality classes as measured at midrapidity in \xexe collisions at \snnXeXe.
        The uncertainties are reported in the order $\pm$ (stat) $\pm$ (syst.).
      }
      \label{tab:dndyXeXe}
      \resizebox{\columnwidth}{!}{%
        \begin{tabular}{lllll}
          \midrule\midrule
          Centrality Class & \avdndypart{\ppis}                           & \avdndypart{\pkas}            & \avdndypart{\pprs}          & \avdndypart{\pphi}                          \\
          \midrule\midrule
          \centint{0}{5}   & 1002.67 $\pm$ 0.39 $\pm$ 57.16               & 149.37 $\pm$ 0.21 $\pm$ 14.07 & 46.21 $\pm$ 0.09 $\pm$ 4.73 & \multirow{2}{*}{9.27 $\pm$ 0.27 $\pm$ 0.95} \\
          \centint{5}{10}  & 808.76 $\pm$ 0.41 $\pm$ 45.34                & 123.58 $\pm$ 0.22 $\pm$ 11.01 & 37.79 $\pm$ 0.09 $\pm$ 3.89 &                                             \\
          \midrule
          \centint{10}{20} & 620.47 $\pm$ 0.24 $\pm$ 34.71                & 95.38 $\pm$ 0.14 $\pm$ 7.94   & 29.26 $\pm$ 0.06 $\pm$ 3.02 & \multirow{2}{*}{5.58 $\pm$ 0.11 $\pm$ 0.64} \\
          \centint{20}{30} & 426.77 $\pm$ 0.21 $\pm$ 24.14                & 66.15 $\pm$ 0.11 $\pm$ 5.44   & 20.74 $\pm$ 0.05 $\pm$ 2.15 &                                             \\
          \midrule
          \centint{30}{40} & 287.20 $\pm$ 0.16 $\pm$ 16.48                & 44.02 $\pm$ 0.09 $\pm$ 3.61   & 14.31 $\pm$ 0.04 $\pm$ 1.48 & \multirow{2}{*}{2.35 $\pm$ 0.07 $\pm$ 0.28} \\
          \centint{40}{50} & 182.89 $\pm$ 0.13 $\pm$ 10.87                & 27.80 $\pm$ 0.07 $\pm$ 2.25   & 9.38 $\pm$ 0.03 $\pm$ 0.98  &                                             \\
          \midrule
          \centint{50}{60} & 111.05 $\pm$ 0.10 $\pm$ 6.62                 & 16.25 $\pm$ 0.05 $\pm$ 1.33   & 5.82 $\pm$ 0.02 $\pm$ 0.61
                           & \multirow{2}{*}{0.84 $\pm$ 0.024 $\pm$ 0.11}                                                                                                             \\
          \centint{60}{70} & 61.23 $\pm$ 0.07 $\pm$ 3.77                  & 8.83 $\pm$ 0.04 $\pm$ 0.77    & 3.26 $\pm$ 0.02 $\pm$ 0.36  &                                             \\
          \midrule
          \centint{70}{90} & 21.43 $\pm$ 0.03 $\pm$ 1.39                  & 2.95 $\pm$ 0.01 $\pm$ 0.26    & 1.17 $\pm$ 0.01 $\pm$ 0.14  & 0.19 $\pm$ 0.01 $\pm$ 0.02                  \\
          \midrule\midrule
        \end{tabular}
      }
    \end{center}
  }
\end{table}

\section{Conclusion and outlook} \label{sec:conclusions}
\label{Sec.:Conclusions}
In this article, results on the \ppipm, \pkapm, \pprp, \pprm and \pphi production measured as a function of centrality in \xexe collisions at \snnXeXe are presented.
For the first time at the LHC, it was possible to disentangle with \AAaa collisions the role of the collision region ``shape'' (eccentricity) and ``size'' (charged-particle multiplicity) on the aspects of the particle production.
The results show a mass dependent enhancement of the particle production at intermediate \pt and a depletion at low \pt.
This feature is more prominent in central collisions and is typically associated with the presence of radial flow.
The effect of the radial flow is reflected in a mass dependent increase of the average momentum for more central collisions.
In light of this interpretation scheme, particles with similar masses receive a similar increase in their average momentum.
This behaviour is confirmed in the comparison of the \meanpt of \ppr and \pphi as a function of \avdNchdeta.
The effect of the radial flow on the production of particles with different masses is investigated by comparing the baryon-to-meson (\rppi and \rpphi) ratios.
A sizable depletion of the low-\pt part of the spectrum is only observed when comparing particles with large mass differences, in agreement with the expectations from the radial flow.
The comparison of particles with similar mass (such as \ppr and \pphi) hints to the fact that the effect is mostly driven by the hadron mass and not by the quark content as one could expect from the recombination of quarks into baryons and mesons.
However, models including recombination of quarks and radial flow are able to reproduce both \rppi and \rpphi at intermediate \pt in central \pbpb collisions suggesting the importance of both mechanisms~\cite{Minissale:2015zwa}.
Moreover, it is found that the results in \xexe and \pbpb collisions are in agreement, indicating that radial flow has a similar magnitude in the two collision systems at LHC energies.
The magnitude of the radial flow is compared in the two systems by using the \rppi ratio in the depletion (1 \gevc) and enhancement (3 \gevc) regions.
It is found that the amount of depletion and enhancement is similar in both cases, while the \vtwo exhibits a clear deviation.
This observation corroborates the intuition that the radial flow depends exclusively on the \avdNchdeta, while anisotropic flow (e.g. \vtwo) depends also on the initial eccentricities of the collision region.

The hadrochemistry is investigated by studying the integrated particle yield ratios of kaons, (anti-)protons, and \pphi to the most abundantly produced pions.
Also, in this case, a behaviour that is mostly driven by \avdNchdeta is observed and thus the intriguing observations from \pbpb collisions related to the \rppi ratio and the \rphipi ratio are now confirmed in a smaller heavy-ion collision system at LHC energies.

As an outlook, these results also pave the way for the future programme of light nuclei collisions at the LHC (in particular the proposed extended future programme with nuclear beams lighter than Pb~\cite{Dainese:2703572}) which is attractive since higher parton luminosities are achievable.
Our results suggest that particle chemistry and radial flow will be driven also in these systems by the final-state charged particle densities.
While \pbpb collisions offer the largest dynamic range in this context, it is also clear from our findings that collisions of small and intermediate nuclei provide an excellent tool to study the hot and strongly-interacting matter in the range of moderate multiplicities.

\newenvironment{acknowledgement}{\relax}{\relax}
\begin{acknowledgement}
  \section*{Acknowledgements}

The ALICE Collaboration would like to thank all its engineers and technicians for their invaluable contributions to the construction of the experiment and the CERN accelerator teams for the outstanding performance of the LHC complex.
The ALICE Collaboration gratefully acknowledges the resources and support provided by all Grid centres and the Worldwide LHC Computing Grid (WLCG) collaboration.
The ALICE Collaboration acknowledges the following funding agencies for their support in building and running the ALICE detector:
A. I. Alikhanyan National Science Laboratory (Yerevan Physics Institute) Foundation (ANSL), State Committee of Science and World Federation of Scientists (WFS), Armenia;
Austrian Academy of Sciences, Austrian Science Fund (FWF): [M 2467-N36] and Nationalstiftung f\"{u}r Forschung, Technologie und Entwicklung, Austria;
Ministry of Communications and High Technologies, National Nuclear Research Center, Azerbaijan;
Conselho Nacional de Desenvolvimento Cient\'{\i}fico e Tecnol\'{o}gico (CNPq), Financiadora de Estudos e Projetos (Finep), Funda\c{c}\~{a}o de Amparo \`{a} Pesquisa do Estado de S\~{a}o Paulo (FAPESP) and Universidade Federal do Rio Grande do Sul (UFRGS), Brazil;
Ministry of Education of China (MOEC) , Ministry of Science \& Technology of China (MSTC) and National Natural Science Foundation of China (NSFC), China;
Ministry of Science and Education and Croatian Science Foundation, Croatia;
Centro de Aplicaciones Tecnol\'{o}gicas y Desarrollo Nuclear (CEADEN), Cubaenerg\'{\i}a, Cuba;
Ministry of Education, Youth and Sports of the Czech Republic, Czech Republic;
The Danish Council for Independent Research | Natural Sciences, the VILLUM FONDEN and Danish National Research Foundation (DNRF), Denmark;
Helsinki Institute of Physics (HIP), Finland;
Commissariat \`{a} l'Energie Atomique (CEA) and Institut National de Physique Nucl\'{e}aire et de Physique des Particules (IN2P3) and Centre National de la Recherche Scientifique (CNRS), France;
Bundesministerium f\"{u}r Bildung und Forschung (BMBF) and GSI Helmholtzzentrum f\"{u}r Schwerionenforschung GmbH, Germany;
General Secretariat for Research and Technology, Ministry of Education, Research and Religions, Greece;
National Research, Development and Innovation Office, Hungary;
Department of Atomic Energy Government of India (DAE), Department of Science and Technology, Government of India (DST), University Grants Commission, Government of India (UGC) and Council of Scientific and Industrial Research (CSIR), India;
Indonesian Institute of Science, Indonesia;
Istituto Nazionale di Fisica Nucleare (INFN), Italy;
Institute for Innovative Science and Technology , Nagasaki Institute of Applied Science (IIST), Japanese Ministry of Education, Culture, Sports, Science and Technology (MEXT) and Japan Society for the Promotion of Science (JSPS) KAKENHI, Japan;
Consejo Nacional de Ciencia (CONACYT) y Tecnolog\'{i}a, through Fondo de Cooperaci\'{o}n Internacional en Ciencia y Tecnolog\'{i}a (FONCICYT) and Direcci\'{o}n General de Asuntos del Personal Academico (DGAPA), Mexico;
Nederlandse Organisatie voor Wetenschappelijk Onderzoek (NWO), Netherlands;
The Research Council of Norway, Norway;
Commission on Science and Technology for Sustainable Development in the South (COMSATS) and Pakistan Atomic Energy Commission, Pakistan;
Pontificia Universidad Cat\'{o}lica del Per\'{u}, Peru;
Ministry of Science and Higher Education, National Science Centre and WUT ID-UB, Poland;
Korea Institute of Science and Technology Information and National Research Foundation of Korea (NRF), Republic of Korea;
Ministry of Education and Scientific Research, Institute of Atomic Physics and Ministry of Research and Innovation and Institute of Atomic Physics, Romania;
Joint Institute for Nuclear Research (JINR), Ministry of Education and Science of the Russian Federation, National Research Centre Kurchatov Institute, Russian Science Foundation and Russian Foundation for Basic Research, Russia;
Ministry of Education, Science, Research and Sport of the Slovak Republic, Slovakia;
National Research Foundation of South Africa, South Africa;
Swedish Research Council (VR) and Knut \& Alice Wallenberg Foundation (KAW), Sweden;
European Organization for Nuclear Research, Switzerland;
Suranaree University of Technology (SUT), National Science and Technology Development Agency (NSDTA) and Office of the Higher Education Commission under NRU project of Thailand, Thailand;
Turkish Atomic Energy Agency (TAEK), Turkey;
National Academy of  Sciences of Ukraine, Ukraine;
Science and Technology Facilities Council (STFC), United Kingdom;
National Science Foundation of the United States of America (NSF) and United States Department of Energy, Office of Nuclear Physics (DOE NP), United States of America.    
\end{acknowledgement}

\bibliographystyle{utphys}   
\bibliography{biblio}

\providecommand{\href}[2]{#2}\begingroup\raggedright\begin{thebibliography}{10}

\bibitem{Abelev:2013haa}
{\bfseries ALICE} Collaboration, B.~Abelev {\em et~al.}, ``{Multiplicity
  Dependence of Pion, Kaon, Proton and Lambda Production in p-Pb Collisions at
  $\sqrt{s_{\rm{NN}}}$ = 5.02 TeV}'',
  \href{http://dx.doi.org/10.1016/j.physletb.2013.11.020}{{\em Phys. Lett.}
  {\bfseries B728} (2014) },
\href{http://arxiv.org/abs/1307.6796}{{\ttfamily arXiv:1307.6796 [nucl-ex]}}.

\bibitem{Abelev:2013vea}
{\bfseries ALICE} Collaboration, B.~Abelev {\em et~al.}, ``{Centrality
  dependence of $\pi$, K, p production in Pb-Pb collisions at
  $\sqrt{s_{\rm{NN}}}$ = 2.76 TeV}'',
  \href{http://dx.doi.org/10.1103/PhysRevC.88.044910}{{\em Phys. Rev.}
  {\bfseries C88} (2013) },
\href{http://arxiv.org/abs/1303.0737}{{\ttfamily arXiv:1303.0737 [hep-ex]}}.

\bibitem{Adam:2015vsf}
{\bfseries ALICE} Collaboration, J.~Adam {\em et~al.}, ``{Multi-strange baryon
  production in p-Pb collisions at $\sqrt{s_{\rm NN}}=5.02$ TeV}'',
  \href{http://dx.doi.org/10.1016/j.physletb.2016.05.027}{{\em Phys. Lett.}
  {\bfseries B758} (2016) },
\href{http://arxiv.org/abs/1512.07227}{{\ttfamily arXiv:1512.07227 [nucl-ex]}}.

\bibitem{Aamodt:2011zza}
{\bfseries ALICE} Collaboration, K.~Aamodt {\em et~al.}, ``{Strange particle
  production in proton-proton collisions at $\sqrt(s) = 0.9 TeV$ with ALICE at
  the LHC}'', \href{http://dx.doi.org/10.1140/epjc/s10052-011-1594-5}{{\em Eur.
  Phys. J.} {\bfseries C71} (2011) },
\href{http://arxiv.org/abs/1012.3257}{{\ttfamily arXiv:1012.3257 [hep-ex]}}.

\bibitem{Aamodt:2011zj}
{\bfseries ALICE} Collaboration, K.~Aamodt {\em et~al.}, ``{Production of
  pions, kaons and protons in $pp$ collisions at $\sqrt{s}= 900$ GeV with ALICE
  at the LHC}'', \href{http://dx.doi.org/10.1140/epjc/s10052-011-1655-9}{{\em
  Eur. Phys. J.} {\bfseries C71} (2011) },
\href{http://arxiv.org/abs/1101.4110}{{\ttfamily arXiv:1101.4110 [hep-ex]}}.

\bibitem{Adam:2016bpr}
{\bfseries ALICE} Collaboration, J.~Adam {\em et~al.}, ``{Production of K$^{*}$
  (892)$^{0}$ and $\phi $ (1020) in p--Pb collisions at $\sqrt{s_{{\text
  {NN}}}}$ = 5.02 TeV}'',
  \href{http://dx.doi.org/10.1140/epjc/s10052-016-4088-7}{{\em Eur. Phys. J. C}
  {\bfseries 76} no.~5, (2016) },
  \href{http://arxiv.org/abs/1601.07868}{{\ttfamily arXiv:1601.07868
  [nucl-ex]}}.

\bibitem{Adam:2017zbf}
{\bfseries ALICE} Collaboration, J.~Adam {\em et~al.}, ``{K$^{*}(892)^{0}$ and
  $\phi(1020)$ meson production at high transverse momentum in pp and Pb-Pb
  collisions at $\sqrt{s_\mathrm{NN}}$ = 2.76 TeV}'',
  \href{http://dx.doi.org/10.1103/PhysRevC.95.064606}{{\em Phys. Rev. C}
  {\bfseries 95} no.~6, (2017) },
  \href{http://arxiv.org/abs/1702.00555}{{\ttfamily arXiv:1702.00555
  [nucl-ex]}}.

\bibitem{ALICE:2017jyt}
{\bfseries ALICE} Collaboration, J.~Adam {\em et~al.}, ``{Enhanced production
  of multi-strange hadrons in high-multiplicity proton-proton collisions}'',
  \href{http://dx.doi.org/10.1038/nphys4111}{{\em Nature Phys.} {\bfseries 13}
  (2017) },
\href{http://arxiv.org/abs/1606.07424}{{\ttfamily arXiv:1606.07424 [nucl-ex]}}.

\bibitem{Acharya:2018orn}
{\bfseries ALICE} Collaboration, S.~Acharya {\em et~al.}, ``{Multiplicity
  dependence of light-flavor hadron production in pp collisions at $\sqrt{s}$ =
  7 TeV}'', \href{http://dx.doi.org/10.1103/PhysRevC.99.024906}{{\em Phys.
  Rev.} {\bfseries C99} no.~2, (2019) },
\href{http://arxiv.org/abs/1807.11321}{{\ttfamily arXiv:1807.11321 [nucl-ex]}}.

\bibitem{Acharya:2019kyh}
{\bfseries ALICE} Collaboration, S.~Acharya {\em et~al.}, ``{Multiplicity
  dependence of (multi-)strange hadron production in proton-proton collisions
  at $\sqrt{s}$ = 13 TeV}'',
  \href{http://dx.doi.org/10.1140/epjc/s10052-020-7673-8}{{\em Eur. Phys. J.}
  {\bfseries C80} no.~2, (2020) },
\href{http://arxiv.org/abs/1908.01861}{{\ttfamily arXiv:1908.01861 [nucl-ex]}}.

\bibitem{Acharya:2019qge}
{\bfseries ALICE} Collaboration, S.~Acharya {\em et~al.}, ``{Evidence of
  rescattering effect in Pb-Pb collisions at the LHC through production of
  $\rm{K}^{*}(892)^{0}$ and $\phi(1020)$ mesons}'',
  \href{http://dx.doi.org/10.1016/j.physletb.2020.135225}{{\em Phys. Lett. B}
  {\bfseries 802} (2020) }, \href{http://arxiv.org/abs/1910.14419}{{\ttfamily
  arXiv:1910.14419 [nucl-ex]}}.

\bibitem{Braun-Munzinger:2015hba}
P.~Braun-Munzinger, V.~Koch, T.~Schäfer, and J.~Stachel, ``{Properties of hot
  and dense matter from relativistic heavy ion collisions}'',
  \href{http://dx.doi.org/10.1016/j.physrep.2015.12.003}{{\em Phys. Rept.}
  {\bfseries 621} (2016) },
\href{http://arxiv.org/abs/1510.00442}{{\ttfamily arXiv:1510.00442 [nucl-th]}}.

\bibitem{Adamczyk:2017iwn}
{\bfseries STAR} Collaboration, L.~Adamczyk {\em et~al.}, ``{Bulk Properties of
  the Medium Produced in Relativistic Heavy-Ion Collisions from the Beam Energy
  Scan Program}'', \href{http://dx.doi.org/10.1103/PhysRevC.96.044904}{{\em
  Phys. Rev.} {\bfseries C96} no.~4, (2017) },
\href{http://arxiv.org/abs/1701.07065}{{\ttfamily arXiv:1701.07065 [nucl-ex]}}.

\bibitem{Kurkela:2018xxd}
A.~Kurkela and A.~Mazeliauskas, ``{Chemical Equilibration in Hadronic
  Collisions}'', \href{http://dx.doi.org/10.1103/PhysRevLett.122.142301}{{\em
  Phys. Rev. Lett.} {\bfseries 122} (2019) },
\href{http://arxiv.org/abs/1811.03040}{{\ttfamily arXiv:1811.03040 [hep-ph]}}.

\bibitem{Bierlich:2018lbp}
C.~Bierlich, ``{Microscopic collectivity: The ridge and strangeness enhancement
  from string–string interactions}'',
  \href{http://dx.doi.org/10.1016/j.nuclphysa.2018.07.015}{{\em Nucl. Phys.}
  {\bfseries A982} (2019) },
\href{http://arxiv.org/abs/1807.05271}{{\ttfamily arXiv:1807.05271 [nucl-th]}}.

\bibitem{Bierlich:2014xba}
C.~Bierlich, G.~Gustafson, L.~Lönnblad, and A.~Tarasov, ``{Effects of
  Overlapping Strings in pp Collisions}'',
  \href{http://dx.doi.org/10.1007/JHEP03(2015)148}{{\em JHEP} {\bfseries 03}
  (2015) },
\href{http://arxiv.org/abs/1412.6259}{{\ttfamily arXiv:1412.6259 [hep-ph]}}.

\bibitem{Abelev:2012wca}
{\bfseries ALICE} Collaboration, B.~Abelev {\em et~al.}, ``{Pion, Kaon, and
  Proton Production in Central Pb--Pb Collisions at $\sqrt{s_{\rm{NN}}} = 2.76$
  TeV}'', \href{http://dx.doi.org/10.1103/PhysRevLett.109.252301}{{\em Phys.
  Rev. Lett.} {\bfseries 109} (2012) },
\href{http://arxiv.org/abs/1208.1974}{{\ttfamily arXiv:1208.1974 [hep-ex]}}.

\bibitem{Noronha-Hostler:2014aia}
J.~Noronha-Hostler and C.~Greiner, ``{Understanding the $p/\pi$ ratio at LHC
  due to QCD mass spectrum}'',
  \href{http://dx.doi.org/10.1016/j.nuclphysa.2014.08.101}{{\em Nucl. Phys.}
  {\bfseries A931} (2014) },
\href{http://arxiv.org/abs/1408.0761}{{\ttfamily arXiv:1408.0761 [nucl-th]}}.

\bibitem{Bellwied:2013cta}
R.~Bellwied, S.~Borsanyi, Z.~Fodor, S.~D. Katz, and C.~Ratti, ``{Is there a
  flavor hierarchy in the deconfinement transition of QCD?}'',
  \href{http://dx.doi.org/10.1103/PhysRevLett.111.202302}{{\em Phys. Rev.
  Lett.} {\bfseries 111} (2013) },
\href{http://arxiv.org/abs/1305.6297}{{\ttfamily arXiv:1305.6297 [hep-lat]}}.

\bibitem{Vovchenko:2018fmh}
V.~Vovchenko, M.~I. Gorenstein, and H.~Stoecker, ``{Finite resonance widths
  influence the thermal-model description of hadron yields}'',
  \href{http://dx.doi.org/10.1103/PhysRevC.98.034906}{{\em Phys. Rev.}
  {\bfseries C98} no.~3, (2018) },
\href{http://arxiv.org/abs/1807.02079}{{\ttfamily arXiv:1807.02079 [nucl-th]}}.

\bibitem{Andronic:2018qqt}
A.~Andronic, P.~Braun-Munzinger, B.~Friman, P.~M. Lo, K.~Redlich, and
  J.~Stachel, ``{The thermal proton yield anomaly in Pb-Pb collisions at the
  LHC and its resolution}'',
  \href{http://dx.doi.org/https://doi.org/10.1016/j.physletb.2019.03.052}{{\em
  Physics Letters B} {\bfseries 792} (2019) },
\href{http://arxiv.org/abs/1808.03102}{{\ttfamily arXiv:1808.03102 [hep-ph]}}.

\bibitem{Vovchenko:2019kes}
V.~Vovchenko, B.~Dönigus, and H.~Stoecker, ``{Canonical statistical model
  analysis of p-p , p-Pb, and Pb-Pb collisions at energies available at the
  CERN Large Hadron Collider}'',
  \href{http://dx.doi.org/10.1103/PhysRevC.100.054906}{{\em Phys. Rev.}
  {\bfseries C100} no.~5, (2019) },
\href{http://arxiv.org/abs/1906.03145}{{\ttfamily arXiv:1906.03145 [hep-ph]}}.

\bibitem{Sharma:2018jqf}
N.~Sharma, J.~Cleymans, B.~Hippolyte, and M.~Paradza, ``{A Comparison of p-p,
  p-Pb, Pb-Pb Collisions in the Thermal Model: Multiplicity Dependence of
  Thermal Parameters}'',
  \href{http://dx.doi.org/10.1103/PhysRevC.99.044914}{{\em Phys. Rev.}
  {\bfseries C99} no.~4, (2019) },
\href{http://arxiv.org/abs/1811.00399}{{\ttfamily arXiv:1811.00399 [hep-ph]}}.

\bibitem{Acharya:2018ihu}
{\bfseries ALICE} Collaboration, S.~Acharya {\em et~al.}, ``{Anisotropic flow
  in Xe-Xe collisions at ${\sqrt{s_{\rm{NN}}} = 5.44}$ TeV}'',
  \href{http://dx.doi.org/10.1016/j.physletb.2018.06.059}{{\em Phys. Lett.}
  {\bfseries B784} (2018) },
\href{http://arxiv.org/abs/1805.01832}{{\ttfamily arXiv:1805.01832 [nucl-ex]}}.

\bibitem{Giacalone:2017dud}
G.~Giacalone, J.~Noronha-Hostler, M.~Luzum, and J.-Y. Ollitrault,
  ``{Hydrodynamic predictions for 5.44 TeV Xe+Xe collisions}'',
  \href{http://dx.doi.org/10.1103/PhysRevC.97.034904}{{\em Phys. Rev.}
  {\bfseries C97} no.~3, (2018) },
\href{http://arxiv.org/abs/1711.08499}{{\ttfamily arXiv:1711.08499 [nucl-th]}}.

\bibitem{PhysRevC.101.044907}
{\bfseries ALICE} Collaboration, S.~Acharya {\em et~al.}, ``Production of
  charged pions, kaons, and (anti-)protons in {Pb-Pb} and inelastic {pp}
  collisions at $\sqrt{{s}_{\rm NN}}=5.02$ {TeV}'',
  \href{http://dx.doi.org/10.1103/PhysRevC.101.044907}{{\em Phys. Rev. C}
  {\bfseries 101} (Apr, 2020) }.
  \url{https://link.aps.org/doi/10.1103/PhysRevC.101.044907}.

\bibitem{Begun:2015ifa}
V.~Begun and W.~Florkowski, ``{Bose-Einstein condensation of pions in heavy-ion
  collisions at the CERN Large Hadron Collider (LHC) energies}'',
  \href{http://dx.doi.org/10.1103/PhysRevC.91.054909}{{\em Phys. Rev.}
  {\bfseries C91} (2015) },
\href{http://arxiv.org/abs/1503.04040}{{\ttfamily arXiv:1503.04040 [nucl-th]}}.

\bibitem{Aamodt:2008zz}
{\bfseries ALICE} Collaboration, K.~Aamodt {\em et~al.}, ``{The ALICE
  experiment at the CERN LHC}'',
\href{http://dx.doi.org/10.1088/1748-0221/3/08/S08002}{{\em JINST} {\bfseries
  3} (2008) }.

\bibitem{Abelev:2014ffa}
{\bfseries ALICE} Collaboration, B.~Abelev {\em et~al.}, ``{Performance of the
  ALICE Experiment at the CERN LHC}'',
  \href{http://dx.doi.org/10.1142/S0217751X14300440}{{\em Int. J. Mod. Phys.}
  {\bfseries A29} (2014) },
\href{http://arxiv.org/abs/1402.4476}{{\ttfamily arXiv:1402.4476 [nucl-ex]}}.

\bibitem{Abbas:2013taa}
{\bfseries ALICE} Collaboration, E.~Abbas {\em et~al.}, ``{Performance of the
  ALICE VZERO system}'',
  \href{http://dx.doi.org/10.1088/1748-0221/8/10/P10016}{{\em JINST} {\bfseries
  8} (2013) },
\href{http://arxiv.org/abs/1306.3130}{{\ttfamily arXiv:1306.3130 [nucl-ex]}}.

\bibitem{Abelev:2013qoq}
{\bfseries ALICE} Collaboration, B.~Abelev {\em et~al.}, ``{Centrality
  determination of Pb-Pb collisions at $\sqrt{s_{\rm{NN}}}$ = 2.76 TeV with
  ALICE}'', \href{http://dx.doi.org/10.1103/PhysRevC.88.044909}{{\em Phys.
  Rev.} {\bfseries C88} no.~4, (2013) },
\href{http://arxiv.org/abs/1301.4361}{{\ttfamily arXiv:1301.4361 [nucl-ex]}}.

\bibitem{ALICE-PUBLIC-2018-003}
{\bfseries ALICE} Collaboration, S.~Acharya {\em et~al.}, ``{Centrality
  determination using the Glauber model in Xe-Xe collisions at $\sqrt{s_{\rm
  NN}} = 5.44$ TeV}'',. \url{https://cds.cern.ch/record/2315401}.
  ALICE-PUBLIC-2018-003.

\bibitem{Acharya:2018hhy}
{\bfseries ALICE} Collaboration, S.~Acharya {\em et~al.}, ``{Centrality and
  pseudorapidity dependence of the charged-particle multiplicity density in
  Xe\textendash{}Xe collisions at $\sqrt{s_{\rm NN}}$ =5.44TeV}'',
  \href{http://dx.doi.org/10.1016/j.physletb.2018.12.048}{{\em Phys. Lett. B}
  {\bfseries 790} (2019) }, \href{http://arxiv.org/abs/1805.04432}{{\ttfamily
  arXiv:1805.04432 [nucl-ex]}}.

\bibitem{Acharya:2018eaq}
{\bfseries ALICE} Collaboration, S.~Acharya {\em et~al.}, ``{Transverse
  momentum spectra and nuclear modification factors of charged particles in
  Xe-Xe collisions at $\sqrt{s_{\rm NN}}$ = 5.44 TeV}'',
  \href{http://dx.doi.org/10.1016/j.physletb.2018.10.052}{{\em Phys. Lett.}
  {\bfseries B788} (2019) },
\href{http://arxiv.org/abs/1805.04399}{{\ttfamily arXiv:1805.04399 [nucl-ex]}}.

\bibitem{Alme:2010ke}
J.~Alme {\em et~al.}, ``{The ALICE TPC, a large 3-dimensional tracking device
  with fast readout for ultra-high multiplicity events}'',
  \href{http://dx.doi.org/10.1016/j.nima.2010.04.042}{{\em Nucl. Instrum.
  Meth.} {\bfseries A622} (2010) },
\href{http://arxiv.org/abs/1001.1950}{{\ttfamily arXiv:1001.1950
  [physics.ins-det]}}.

\bibitem{Akindinov:2013tea}
A.~Akindinov {\em et~al.}, ``{Performance of the ALICE Time-Of-Flight detector
  at the LHC}'',
\href{http://dx.doi.org/10.1140/epjp/i2013-13044-x}{{\em Eur. Phys. J. Plus}
  {\bfseries 128} (2013) }.

\bibitem{Beole:1998yq}
{\bfseries ALICE} Collaboration, S.~Beole {\em et~al.}, ``{ALICE technical
  design report: Detector for high momentum PID}'',.
\url{https://cds.cern.ch/record/381431}.

\bibitem{ALICE-PUBLIC-2017-005}
{\bfseries ALICE} Collaboration, S.~Acharya {\em et~al.}, ``{The ALICE
  definition of primary particles}'',.
  \url{https://cds.cern.ch/record/2270008}.

\bibitem{Abelevetal:2014dna}
{\bfseries ALICE} Collaboration, B.~Abelev {\em et~al.}, ``{Technical Design
  Report for the Upgrade of the ALICE Inner Tracking System}'',
\href{http://dx.doi.org/10.1088/0954-3899/41/8/087002}{{\em J. Phys.}
  {\bfseries G41} (2014) }.

\bibitem{DAgostini:1994fjx}
G.~D'Agostini, ``{A Multidimensional unfolding method based on Bayes'
  theorem}'', \href{http://dx.doi.org/10.1016/0168-9002(95)00274-X}{{\em Nucl.
  Instrum. Meth. A} {\bfseries 362} (1995) }.

\bibitem{Wang:1991hta}
X.-N. Wang and M.~Gyulassy, ``{HIJING: A Monte Carlo model for multiple jet
  production in p p, p A and A A collisions}'',
\href{http://dx.doi.org/10.1103/PhysRevD.44.3501}{{\em Phys. Rev.} {\bfseries
  D44} (1991) }.

\bibitem{Adam:2015qaa}
{\bfseries ALICE} Collaboration, J.~Adam {\em et~al.}, ``{Measurement of pion,
  kaon and proton production in proton$-$proton collisions at $\sqrt{s} = 7$
  TeV}'', \href{http://dx.doi.org/10.1140/epjc/s10052-015-3422-9}{{\em Eur.
  Phys. J.} {\bfseries C75} no.~5, (2015) },
\href{http://arxiv.org/abs/1504.00024}{{\ttfamily arXiv:1504.00024 [nucl-ex]}}.

\bibitem{Zyla:2020zbs}
{\bfseries Particle Data Group} Collaboration, P.~Zyla {\em et~al.}, ``{Review
  of Particle Physics}'', \href{http://dx.doi.org/10.1093/ptep/ptaa104}{{\em
  PTEP} {\bfseries 2020} no.~8, (2020) }.

\bibitem{Brun:1119728}
R.~Brun, F.~Bruyant, M.~Maire, A.~C. McPherson, and P.~Zanarini, {\em {GEANT 3:
  user's guide Geant 3.10, Geant 3.11; rev. version}}.
\newblock CERN, Geneva, 1987.
\newblock \url{https://cds.cern.ch/record/1119728}.

\bibitem{Aamodt:2010dx}
{\bfseries ALICE} Collaboration, K.~Aamodt {\em et~al.}, ``{Midrapidity
  antiproton-to-proton ratio in pp collisions at $\sqrt{s} = 0.9$ and $7$~TeV
  measured by the ALICE experiment}'',
  \href{http://dx.doi.org/10.1103/PhysRevLett.105.072002}{{\em Phys. Rev.
  Lett.} {\bfseries 105} (2010) },
\href{http://arxiv.org/abs/1006.5432}{{\ttfamily arXiv:1006.5432 [hep-ex]}}.

\bibitem{Agostinelli:2002hh}
{\bfseries GEANT4} Collaboration, S.~Agostinelli {\em et~al.}, ``{GEANT4: A
  Simulation toolkit}'',
\href{http://dx.doi.org/10.1016/S0168-9002(03)01368-8}{{\em Nucl. Instrum.
  Meth.} {\bfseries A506} (2003) }.

\bibitem{Battistoni:2007zzb}
G.~Battistoni, S.~Muraro, P.~R. Sala, F.~Cerutti, A.~Ferrari, S.~Roesler,
  A.~Fasso, and J.~Ranft, ``{The FLUKA code: Description and benchmarking}'',
  \href{http://dx.doi.org/10.1063/1.2720455}{{\em AIP Conf. Proc.} {\bfseries
  896} (2007) }.
[,31(2007)].

\bibitem{Abelev:2014uua}
{\bfseries ALICE} Collaboration, B.~Abelev {\em et~al.}, ``{$K^*(892)^0$ and
  $\phi(1020)$ production in Pb-Pb collisions at $\sqrt{s_{\rm NN}}$ = 2.76
  TeV}'', \href{http://dx.doi.org/10.1103/PhysRevC.91.024609}{{\em Phys. Rev.
  C} {\bfseries 91} (2015) }, \href{http://arxiv.org/abs/1404.0495}{{\ttfamily
  arXiv:1404.0495 [nucl-ex]}}.

\bibitem{Abelev:2012cn}
{\bfseries ALICE} Collaboration, B.~Abelev {\em et~al.}, ``{Neutral pion and
  $\eta$ meson production in proton-proton collisions at $\sqrt{s}=0.9$ TeV and
  $\sqrt{s}=7$ TeV}'',
  \href{http://dx.doi.org/10.1016/j.physletb.2012.09.015}{{\em Phys. Lett. B}
  {\bfseries 717} (2012) }, \href{http://arxiv.org/abs/1205.5724}{{\ttfamily
  arXiv:1205.5724 [hep-ex]}}.

\bibitem{Abbas:2013rua}
{\bfseries ALICE} Collaboration, E.~Abbas {\em et~al.}, ``{Mid-rapidity
  anti-baryon to baryon ratios in pp collisions at $\sqrt{s}$ = 0.9, 2.76 and 7
  TeV measured by ALICE}'',
  \href{http://dx.doi.org/10.1140/epjc/s10052-013-2496-5}{{\em Eur. Phys. J. C}
  {\bfseries 73} (2013) }, \href{http://arxiv.org/abs/1305.1562}{{\ttfamily
  arXiv:1305.1562 [nucl-ex]}}.

\bibitem{Schnedermann:1993ws}
E.~Schnedermann, J.~Sollfrank, and U.~W. Heinz, ``{Thermal phenomenology of
  hadrons from 200 AGeV S+S collisions}'',
  \href{http://dx.doi.org/10.1103/PhysRevC.48.2462}{{\em Phys. Rev.} {\bfseries
  C48} (1993) },
\href{http://arxiv.org/abs/nucl-th/9307020}{{\ttfamily arXiv:nucl-th/9307020
  [nucl-th]}}.

\bibitem{Greco:2003xt}
V.~Greco, C.~M. Ko, and P.~Levai, ``{Parton coalescence and anti-proton / pion
  anomaly at RHIC}'',
  \href{http://dx.doi.org/10.1103/PhysRevLett.90.202302}{{\em Phys. Rev. Lett.}
  {\bfseries 90} (2003) },
\href{http://arxiv.org/abs/nucl-th/0301093}{{\ttfamily arXiv:nucl-th/0301093
  [nucl-th]}}.

\bibitem{Fries:2003vb}
R.~J. Fries, B.~M{\"u}ller, C.~Nonaka, and S.~A. Bass, ``{Hadronization in
  heavy ion collisions: Recombination and fragmentation of partons}'',
  \href{http://dx.doi.org/10.1103/PhysRevLett.90.202303}{{\em Phys. Rev. Lett.}
  {\bfseries 90} (2003) },
\href{http://arxiv.org/abs/nucl-th/0301087}{{\ttfamily arXiv:nucl-th/0301087
  [nucl-th]}}.

\bibitem{Minissale:2015zwa}
V.~Minissale, F.~Scardina, and V.~Greco, ``{Hadrons from coalescence plus
  fragmentation in AA collisions at energies available at the BNL Relativistic
  Heavy Ion Collider to the CERN Large Hadron Collider}'',
  \href{http://dx.doi.org/10.1103/PhysRevC.92.054904}{{\em Phys. Rev.}
  {\bfseries C92} no.~5, (2015) },
\href{http://arxiv.org/abs/1502.06213}{{\ttfamily arXiv:1502.06213 [nucl-th]}}.

\bibitem{Acharya:2018zuq}
{\bfseries ALICE} Collaboration, S.~Acharya {\em et~al.}, ``{Anisotropic flow
  of identified particles in Pb-Pb collisions at $
  {\sqrt{s}}_{\mathrm{NN}}=5.02 $ TeV}'',
  \href{http://dx.doi.org/10.1007/JHEP09(2018)006}{{\em JHEP} {\bfseries 09}
  (2018) },
\href{http://arxiv.org/abs/1805.04390}{{\ttfamily arXiv:1805.04390 [nucl-ex]}}.

\bibitem{Cleymans:2020fsc}
J.~Cleymans, P.~M. Lo, K.~Redlich, and N.~Sharma, ``{Multiplicity dependence of
  (multi)strange baryons in the canonical ensemble with phase shift
  corrections}'', \href{http://dx.doi.org/10.1103/PhysRevC.103.014904}{{\em
  Phys. Rev. C} {\bfseries 103} no.~1, (2021) },
  \href{http://arxiv.org/abs/2009.04844}{{\ttfamily arXiv:2009.04844
  [hep-ph]}}.

\bibitem{Becattini:2014hla}
F.~Becattini, E.~Grossi, M.~Bleicher, J.~Steinheimer, and R.~Stock,
  ``{Centrality dependence of hadronization and chemical freeze-out conditions
  in heavy ion collisions at $\sqrt s_{\rm{NN}}$ = 2.76 TeV}'',
  \href{http://dx.doi.org/10.1103/PhysRevC.90.054907}{{\em Phys. Rev.}
  {\bfseries C90} no.~5, (2014) },
\href{http://arxiv.org/abs/1405.0710}{{\ttfamily arXiv:1405.0710 [nucl-th]}}.

\bibitem{Acharya:2019bli}
{\bfseries ALICE} Collaboration, S.~Acharya {\em et~al.}, ``{Multiplicity
  dependence of K*(892)$^{0}$ and $\phi$(1020) production in pp collisions at
  $\sqrt {s}$ =13 TeV}'',
  \href{http://dx.doi.org/10.1016/j.physletb.2020.135501}{{\em Phys. Lett. B}
  {\bfseries 807} (2020) }, \href{http://arxiv.org/abs/1910.14397}{{\ttfamily
  arXiv:1910.14397 [nucl-ex]}}.

\bibitem{Vovchenko:privatecomm}
V.~Vovchenko and B.~Dönigus. {Private Communication}, 2021.

\bibitem{Dainese:2703572}
A.~Dainese, M.~Mangano, A.~B. Meyer, A.~Nisati, G.~Salam, and M.~A. Vesterinen,
  \href{http://dx.doi.org/10.23731/CYRM-2019-007}{``{Report on the Physics at
  the HL-LHC, and Perspectives for the HE-LHC}'',} Tech. Rep. CERN-2019-007,
  Geneva, Switzerland, 2019.
\newblock \url{http://cds.cern.ch/record/2703572}.

\end{thebibliography}\endgroup

\newpage
\appendix
\section{The ALICE Collaboration}
\label{app:collab}
\begingroup
\small
\begin{flushleft} 


S.~Acharya$^{\rm 143}$, 
D.~Adamov\'{a}$^{\rm 97}$, 
A.~Adler$^{\rm 75}$, 
J.~Adolfsson$^{\rm 82}$, 
G.~Aglieri Rinella$^{\rm 35}$, 
M.~Agnello$^{\rm 31}$, 
N.~Agrawal$^{\rm 55}$, 
Z.~Ahammed$^{\rm 143}$, 
S.~Ahmad$^{\rm 16}$, 
S.U.~Ahn$^{\rm 77}$
% ...


\section*{Affiliation notes}


$^{\rm I}$ Deceased\\
$^{\rm II}$ Also at: Italian National Agency for New Technologies, Energy and Sustainable Economic Development (ENEA), Bologna, Italy\\
$^{\rm III}$ Also at: Dipartimento DET del Politecnico di Torino, Turin, Italy\\
$^{\rm IV}$ Also at: M.V. Lomonosov Moscow State University, D.V. Skobeltsyn Institute of Nuclear, Physics, Moscow, Russia\\
$^{\rm V}$ Also at: Institute of Theoretical Physics, University of Wroclaw, Poland\\



\section*{Collaboration Institutes}


$^{1}$ A.I. Alikhanyan National Science Laboratory (Yerevan Physics Institute) Foundation, Yerevan, Armenia\\
$^{2}$ AGH University of Science and Technology, Cracow, Poland\\
$^{3}$ Bogolyubov Institute for Theoretical Physics, National Academy of Sciences of Ukraine, Kiev, Ukraine\\
$^{4}$ Bose Institute, Department of Physics  and Centre for Astroparticle Physics and Space Science (CAPSS), Kolkata, India\\
$^{5}$ Budker Institute for Nuclear Physics, Novosibirsk, Russia\\
$^{6}$ California Polytechnic State University, San Luis Obispo, California, United States\\
$^{7}$ Central China Normal University, Wuhan, China\\
$^{8}$ Centro de Aplicaciones Tecnol\'{o}gicas y Desarrollo Nuclear (CEADEN), Havana, Cuba\\
$^{9}$ Centro de Investigaci\'{o}n y de Estudios Avanzados (CINVESTAV), Mexico City and M\'{e}rida, Mexico\\
$^{10}$ Chicago State University, Chicago, Illinois, United States\\
$^{11}$ China Institute of Atomic Energy, Beijing, China\\
$^{12}$ Chungbuk National University, Cheongju, Republic of Korea\\
$^{13}$ Comenius University Bratislava, Faculty of Mathematics, Physics and Informatics, Bratislava, Slovakia\\
$^{14}$ COMSATS University Islamabad, Islamabad, Pakistan\\
$^{15}$ Creighton University, Omaha, Nebraska, United States\\
$^{16}$ Department of Physics, Aligarh Muslim University, Aligarh, India\\
$^{17}$ Department of Physics, Pusan National University, Pusan, Republic of Korea\\
$^{18}$ Department of Physics, Sejong University, Seoul, Republic of Korea\\
$^{19}$ Department of Physics, University of California, Berkeley, California, United States\\
$^{20}$ Department of Physics, University of Oslo, Oslo, Norway\\
$^{21}$ Department of Physics and Technology, University of Bergen, Bergen, Norway\\
$^{22}$ Dipartimento di Fisica dell'Universit\`{a} 'La Sapienza' and Sezione INFN, Rome, Italy\\
$^{23}$ Dipartimento di Fisica dell'Universit\`{a} and Sezione INFN, Cagliari, Italy\\
$^{24}$ Dipartimento di Fisica dell'Universit\`{a} and Sezione INFN, Trieste, Italy\\
$^{25}$ Dipartimento di Fisica dell'Universit\`{a} and Sezione INFN, Turin, Italy\\
$^{26}$ Dipartimento di Fisica e Astronomia dell'Universit\`{a} and Sezione INFN, Bologna, Italy\\
$^{27}$ Dipartimento di Fisica e Astronomia dell'Universit\`{a} and Sezione INFN, Catania, Italy\\
$^{28}$ Dipartimento di Fisica e Astronomia dell'Universit\`{a} and Sezione INFN, Padova, Italy\\
$^{29}$ Dipartimento di Fisica e Nucleare e Teorica, Universit\`{a} di Pavia  and Sezione INFN, Pavia, Italy\\
$^{30}$ Dipartimento di Fisica `E.R.~Caianiello' dell'Universit\`{a} and Gruppo Collegato INFN, Salerno, Italy\\
$^{31}$ Dipartimento DISAT del Politecnico and Sezione INFN, Turin, Italy\\
$^{32}$ Dipartimento di Scienze e Innovazione Tecnologica dell'Universit\`{a} del Piemonte Orientale and INFN Sezione di Torino, Alessandria, Italy\\
$^{33}$ Dipartimento di Scienze MIFT, Universit\`{a} di Messina, Messina, Italy\\
$^{34}$ Dipartimento Interateneo di Fisica `M.~Merlin' and Sezione INFN, Bari, Italy\\
$^{35}$ European Organization for Nuclear Research (CERN), Geneva, Switzerland\\
$^{36}$ Faculty of Electrical Engineering, Mechanical Engineering and Naval Architecture, University of Split, Split, Croatia\\
$^{37}$ Faculty of Engineering and Science, Western Norway University of Applied Sciences, Bergen, Norway\\
$^{38}$ Faculty of Nuclear Sciences and Physical Engineering, Czech Technical University in Prague, Prague, Czech Republic\\
$^{39}$ Faculty of Science, P.J.~\v{S}af\'{a}rik University, Ko\v{s}ice, Slovakia\\
$^{40}$ Frankfurt Institute for Advanced Studies, Johann Wolfgang Goethe-Universit\"{a}t Frankfurt, Frankfurt, Germany\\
$^{41}$ Fudan University, Shanghai, China\\
$^{42}$ Gangneung-Wonju National University, Gangneung, Republic of Korea\\
$^{43}$ Gauhati University, Department of Physics, Guwahati, India\\
$^{44}$ Helmholtz-Institut f\"{u}r Strahlen- und Kernphysik, Rheinische Friedrich-Wilhelms-Universit\"{a}t Bonn, Bonn, Germany\\
$^{45}$ Helsinki Institute of Physics (HIP), Helsinki, Finland\\
$^{46}$ High Energy Physics Group,  Universidad Aut\'{o}noma de Puebla, Puebla, Mexico\\
$^{47}$ Hiroshima University, Hiroshima, Japan\\
$^{48}$ Hochschule Worms, Zentrum  f\"{u}r Technologietransfer und Telekommunikation (ZTT), Worms, Germany\\
$^{49}$ Horia Hulubei National Institute of Physics and Nuclear Engineering, Bucharest, Romania\\
$^{50}$ Indian Institute of Technology Bombay (IIT), Mumbai, India\\
$^{51}$ Indian Institute of Technology Indore, Indore, India\\
$^{52}$ Indonesian Institute of Sciences, Jakarta, Indonesia\\
$^{53}$ INFN, Laboratori Nazionali di Frascati, Frascati, Italy\\
$^{54}$ INFN, Sezione di Bari, Bari, Italy\\
$^{55}$ INFN, Sezione di Bologna, Bologna, Italy\\
$^{56}$ INFN, Sezione di Cagliari, Cagliari, Italy\\
$^{57}$ INFN, Sezione di Catania, Catania, Italy\\
$^{58}$ INFN, Sezione di Padova, Padova, Italy\\
$^{59}$ INFN, Sezione di Roma, Rome, Italy\\
$^{60}$ INFN, Sezione di Torino, Turin, Italy\\
$^{61}$ INFN, Sezione di Trieste, Trieste, Italy\\
$^{62}$ Inha University, Incheon, Republic of Korea\\
$^{63}$ Institute for Gravitational and Subatomic Physics (GRASP), Utrecht University/Nikhef, Utrecht, Netherlands\\
$^{64}$ Institute for Nuclear Research, Academy of Sciences, Moscow, Russia\\
$^{65}$ Institute of Experimental Physics, Slovak Academy of Sciences, Ko\v{s}ice, Slovakia\\
$^{66}$ Institute of Physics, Homi Bhabha National Institute, Bhubaneswar, India\\
$^{67}$ Institute of Physics of the Czech Academy of Sciences, Prague, Czech Republic\\
$^{68}$ Institute of Space Science (ISS), Bucharest, Romania\\
$^{69}$ Institut f\"{u}r Kernphysik, Johann Wolfgang Goethe-Universit\"{a}t Frankfurt, Frankfurt, Germany\\
$^{70}$ Instituto de Ciencias Nucleares, Universidad Nacional Aut\'{o}noma de M\'{e}xico, Mexico City, Mexico\\
$^{71}$ Instituto de F\'{i}sica, Universidade Federal do Rio Grande do Sul (UFRGS), Porto Alegre, Brazil\\
$^{72}$ Instituto de F\'{\i}sica, Universidad Nacional Aut\'{o}noma de M\'{e}xico, Mexico City, Mexico\\
$^{73}$ iThemba LABS, National Research Foundation, Somerset West, South Africa\\
$^{74}$ Jeonbuk National University, Jeonju, Republic of Korea\\
$^{75}$ Johann-Wolfgang-Goethe Universit\"{a}t Frankfurt Institut f\"{u}r Informatik, Fachbereich Informatik und Mathematik, Frankfurt, Germany\\
$^{76}$ Joint Institute for Nuclear Research (JINR), Dubna, Russia\\
$^{77}$ Korea Institute of Science and Technology Information, Daejeon, Republic of Korea\\
$^{78}$ KTO Karatay University, Konya, Turkey\\
$^{79}$ Laboratoire de Physique des 2 Infinis, Ir\`{e}ne Joliot-Curie, Orsay, France\\
$^{80}$ Laboratoire de Physique Subatomique et de Cosmologie, Universit\'{e} Grenoble-Alpes, CNRS-IN2P3, Grenoble, France\\
$^{81}$ Lawrence Berkeley National Laboratory, Berkeley, California, United States\\
$^{82}$ Lund University Department of Physics, Division of Particle Physics, Lund, Sweden\\
$^{83}$ Moscow Institute for Physics and Technology, Moscow, Russia\\
$^{84}$ Nagasaki Institute of Applied Science, Nagasaki, Japan\\
$^{85}$ Nara Women{'}s University (NWU), Nara, Japan\\
$^{86}$ National and Kapodistrian University of Athens, School of Science, Department of Physics , Athens, Greece\\
$^{87}$ National Centre for Nuclear Research, Warsaw, Poland\\
$^{88}$ National Institute of Science Education and Research, Homi Bhabha National Institute, Jatni, India\\
$^{89}$ National Nuclear Research Center, Baku, Azerbaijan\\
$^{90}$ National Research Centre Kurchatov Institute, Moscow, Russia\\
$^{91}$ Niels Bohr Institute, University of Copenhagen, Copenhagen, Denmark\\
$^{92}$ Nikhef, National institute for subatomic physics, Amsterdam, Netherlands\\
$^{93}$ NRC Kurchatov Institute IHEP, Protvino, Russia\\
$^{94}$ NRC \guillemotleft Kurchatov\guillemotright  Institute - ITEP, Moscow, Russia\\
$^{95}$ NRNU Moscow Engineering Physics Institute, Moscow, Russia\\
$^{96}$ Nuclear Physics Group, STFC Daresbury Laboratory, Daresbury, United Kingdom\\
$^{97}$ Nuclear Physics Institute of the Czech Academy of Sciences, \v{R}e\v{z} u Prahy, Czech Republic\\
$^{98}$ Oak Ridge National Laboratory, Oak Ridge, Tennessee, United States\\
$^{99}$ Ohio State University, Columbus, Ohio, United States\\
$^{100}$ Petersburg Nuclear Physics Institute, Gatchina, Russia\\
$^{101}$ Physics department, Faculty of science, University of Zagreb, Zagreb, Croatia\\
$^{102}$ Physics Department, Panjab University, Chandigarh, India\\
$^{103}$ Physics Department, University of Jammu, Jammu, India\\
$^{104}$ Physics Department, University of Rajasthan, Jaipur, India\\
$^{105}$ Physikalisches Institut, Eberhard-Karls-Universit\"{a}t T\"{u}bingen, T\"{u}bingen, Germany\\
$^{106}$ Physikalisches Institut, Ruprecht-Karls-Universit\"{a}t Heidelberg, Heidelberg, Germany\\
$^{107}$ Physik Department, Technische Universit\"{a}t M\"{u}nchen, Munich, Germany\\
$^{108}$ PINSTECH, Islamabad, Pakistan\\
$^{109}$ Politecnico di Bari and Sezione INFN, Bari, Italy\\
$^{110}$ Research Division and ExtreMe Matter Institute EMMI, GSI Helmholtzzentrum f\"ur Schwerionenforschung GmbH, Darmstadt, Germany\\
$^{111}$ Rudjer Bo\v{s}kovi\'{c} Institute, Zagreb, Croatia\\
$^{112}$ Russian Federal Nuclear Center (VNIIEF), Sarov, Russia\\
$^{113}$ Saha Institute of Nuclear Physics, Homi Bhabha National Institute, Kolkata, India\\
$^{114}$ School of Physics and Astronomy, University of Birmingham, Birmingham, United Kingdom\\
$^{115}$ Secci\'{o}n F\'{\i}sica, Departamento de Ciencias, Pontificia Universidad Cat\'{o}lica del Per\'{u}, Lima, Peru\\
$^{116}$ St. Petersburg State University, St. Petersburg, Russia\\
$^{117}$ Stefan Meyer Institut f\"{u}r Subatomare Physik (SMI), Vienna, Austria\\
$^{118}$ SUBATECH, IMT Atlantique, Universit\'{e} de Nantes, CNRS-IN2P3, Nantes, France\\
$^{119}$ Suranaree University of Technology, Nakhon Ratchasima, Thailand\\
$^{120}$ Technical University of Ko\v{s}ice, Ko\v{s}ice, Slovakia\\
$^{121}$ The Henryk Niewodniczanski Institute of Nuclear Physics, Polish Academy of Sciences, Cracow, Poland\\
$^{122}$ The University of Texas at Austin, Austin, Texas, United States\\
$^{123}$ Universidad Aut\'{o}noma de Sinaloa, Culiac\'{a}n, Mexico\\
$^{124}$ Universidade de S\~{a}o Paulo (USP), S\~{a}o Paulo, Brazil\\
$^{125}$ Universidade Estadual de Campinas (UNICAMP), Campinas, Brazil\\
$^{126}$ Universidade Federal do ABC, Santo Andre, Brazil\\
$^{127}$ University of Cape Town, Cape Town, South Africa\\
$^{128}$ University of Houston, Houston, Texas, United States\\
$^{129}$ University of Jyv\"{a}skyl\"{a}, Jyv\"{a}skyl\"{a}, Finland\\
$^{130}$ University of Liverpool, Liverpool, United Kingdom\\
$^{131}$ University of Science and Technology of China, Hefei, China\\
$^{132}$ University of South-Eastern Norway, Tonsberg, Norway\\
$^{133}$ University of Tennessee, Knoxville, Tennessee, United States\\
$^{134}$ University of the Witwatersrand, Johannesburg, South Africa\\
$^{135}$ University of Tokyo, Tokyo, Japan\\
$^{136}$ University of Tsukuba, Tsukuba, Japan\\
$^{137}$ Universit\'{e} Clermont Auvergne, CNRS/IN2P3, LPC, Clermont-Ferrand, France\\
$^{138}$ Universit\'{e} de Lyon, CNRS/IN2P3, Institut de Physique des 2 Infinis de Lyon , Lyon, France\\
$^{139}$ Universit\'{e} de Strasbourg, CNRS, IPHC UMR 7178, F-67000 Strasbourg, France, Strasbourg, France\\
$^{140}$ Universit\'{e} Paris-Saclay Centre d'Etudes de Saclay (CEA), IRFU, D\'{e}partment de Physique Nucl\'{e}aire (DPhN), Saclay, France\\
$^{141}$ Universit\`{a} degli Studi di Foggia, Foggia, Italy\\
$^{142}$ Universit\`{a} di Brescia and Sezione INFN, Brescia, Italy\\
$^{143}$ Variable Energy Cyclotron Centre, Homi Bhabha National Institute, Kolkata, India\\
$^{144}$ Warsaw University of Technology, Warsaw, Poland\\
$^{145}$ Wayne State University, Detroit, Michigan, United States\\
$^{146}$ Westf\"{a}lische Wilhelms-Universit\"{a}t M\"{u}nster, Institut f\"{u}r Kernphysik, M\"{u}nster, Germany\\
$^{147}$ Wigner Research Centre for Physics, Budapest, Hungary\\
$^{148}$ Yale University, New Haven, Connecticut, United States\\
$^{149}$ Yonsei University, Seoul, Republic of Korea\\

\bigskip 

\end{flushleft} 
\endgroup

\end{document}